\documentclass[11pt,a4paper]{article}
\usepackage[dvipdfmx]{graphicx}
\usepackage{jheppub}
\usepackage{amsfonts,amsmath,amssymb,epsf}
\usepackage{amsmath,braket,mathtools}
\usepackage{graphicx,hyperref,color}
\usepackage{subcaption} 
\usepackage{refcount}
\usepackage{tikz}
\usetikzlibrary{decorations.pathmorphing,shapes}
\usetikzlibrary {arrows.meta}
\hypersetup{
    bookmarks=true,         
    unicode=false,          
    pdftoolbar=true,        
    pdfmenubar=true,        
    linktocpage=true,       
    pdffitwindow=false,     
    pdfstartview={FitH},    
    pdfnewwindow=true,      
    colorlinks=true,       
    linkcolor=blue,          
    citecolor=blue,        
    filecolor=blue,      
    urlcolor=blue           
}

\newcommand{\be}{\begin{equation}}
\newcommand{\ba}{\begin{eqnarray}}
\newcommand{\ea}{\end{eqnarray}}
\newcommand{\ee}{\end{equation}}

\newcommand{\s}{\sqrt}

\newcommand{\ap}{\alpha}

\newcommand{\no}{\nonumber \\}
\newcommand{\la}{\langle}
\newcommand{\lb}{\rangle}
\newcommand{\bea}{\begin{eqnarray}}
\newcommand{\eea}{\end{eqnarray}}
\newcommand{\bes}{\begin{equation*}}
\newcommand{\beas}{\begin{eqnarray*}}
\newcommand{\eeas}{\end{eqnarray*}}
\newcommand{\bas}{\begin{array*}}
\newcommand{\eas}{\end{array*}}
\newcommand{\ees}{\end{equation*}}
\newcommand{\nn}{\nonumber}

\newcommand{\ep}{\epsilon}

\renewcommand{\Re}{\operatorname{Re}}
\renewcommand{\Im}{\operatorname{Im}}

\subheader{\today}
\title{\boldmath Holographic Entanglement Propagation Through Wormholes}

\author[a]{Kazuki Doi,}
\author[a]{Liang Li,}
\author[a]{Ung Nguyen,}
\author[a,b]{Tadashi Takayanagi}
\affiliation[a]{Center for Gravitational Physics and Quantum Information,\\ 
Yukawa Institute for Theoretical Physics, Kyoto University,\\ 
Kitashirakawa Oiwakecho, Sakyo-ku, Kyoto 606-8502, Japan}
\affiliation[b]{Inamori Research Institute for Science,\\
620 Suiginya-cho, Shimogyo-ku, Kyoto 600-8411 Japan}

\abstract{We study how energy and quantum entanglement are transferred when two identical CFTs are entangled locally. This is probed by considering a local operator insertion in one of the CFTs. When the CFTs have holographic duals via the AdS/CFT correspondence, the transfer happens through an AdS wormhole that allows signal propagation even beyond the horizon from one AdS boundary to the other; we demonstrate this in explicit CFT calculations. We argue that this transmission is possible because the insertion of a local operator is not a unitary process but a regularized version of projection measurement, and that this is interpreted as quantum teleportation. We also find that this leads to a phenomenon opposite to scrambling, where mutual information, instead of being suppressed, gets enhanced by the insertion of a local operator excitation.}

\begin{document} 

\begin{flushright}
YITP-26-09\\
\end{flushright}
\maketitle
\flushbottom

\clearpage
\section{Introduction}\label{Sec:Intro}

Quantum entanglement has played a crucial role in high energy theory as it provides new methods to analyze quantum field theories and quantum gravity. For example, the measure of quantum entanglement for pure states, known as entanglement entropy, provides a universal characterization of degrees of freedom in quantum many-body systems and field theories \cite{Vidal:2002rm,HLW,CCR,CH,Ni}. Moreover, the entanglement entropy has a direct geometric interpretation via the holographic entanglement entropy \cite{RT,HRT} in light of the AdS/CFT correspondence \cite{Maldacena:1997re,Gubser:1998bc,Witten:1998qj}.

In spite of these developments, we still do not seem to incorporate the full idea of quantum entanglement into our studies. One major reason for this is that in quantum information theory, interesting physical processes are described by quantum operations, most notably local operations and classical communications (LOCC) \cite{Book,Geo,Wilde}. For example, one key feature of quantum entanglement is that it does not increase under LOCCs \cite{Bennett:1995tk,Nielsen:1999zza}. Various important processes in quantum information theory such as quantum teleportation are described as examples of LOCC. 
On the other hand, there have only been a very limited number of works in high energy theory which sought to take advantage of such operational aspects of quantum entanglement. They include realizations of projection measurement in conformal field theories (CFTs) \cite{Raj,Rajj,Rajjj} and construction of their holographic duals \cite{Numasawa:2016emc} as well as the holographic description of traversable wormholes \cite{Gao:2016bin,Maldacena:2018lmt}, most of which are closely related to quantum teleportation.

\begin{figure}
    \centering
    \includegraphics[width=7.5cm]{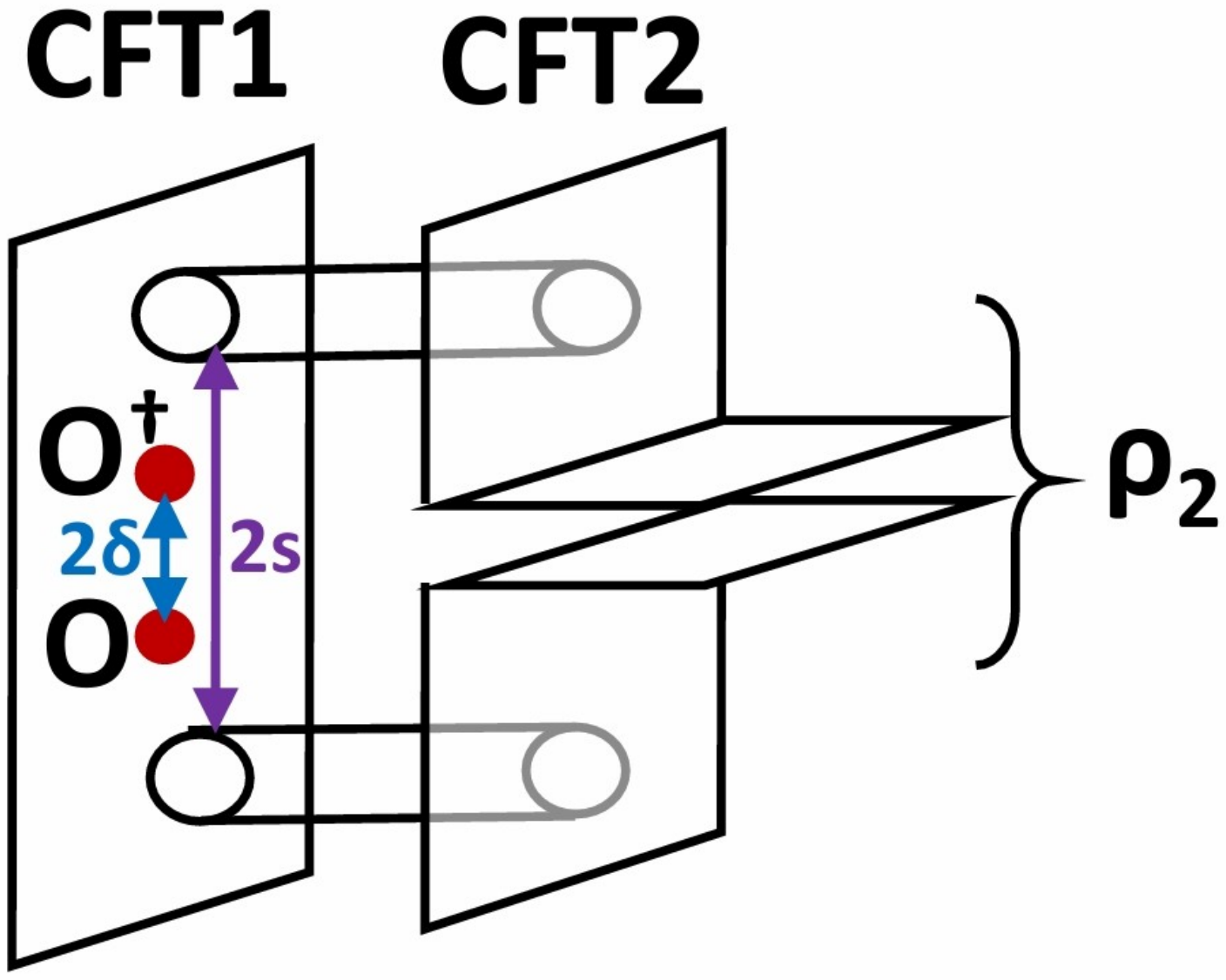}
     \includegraphics[width=7cm]{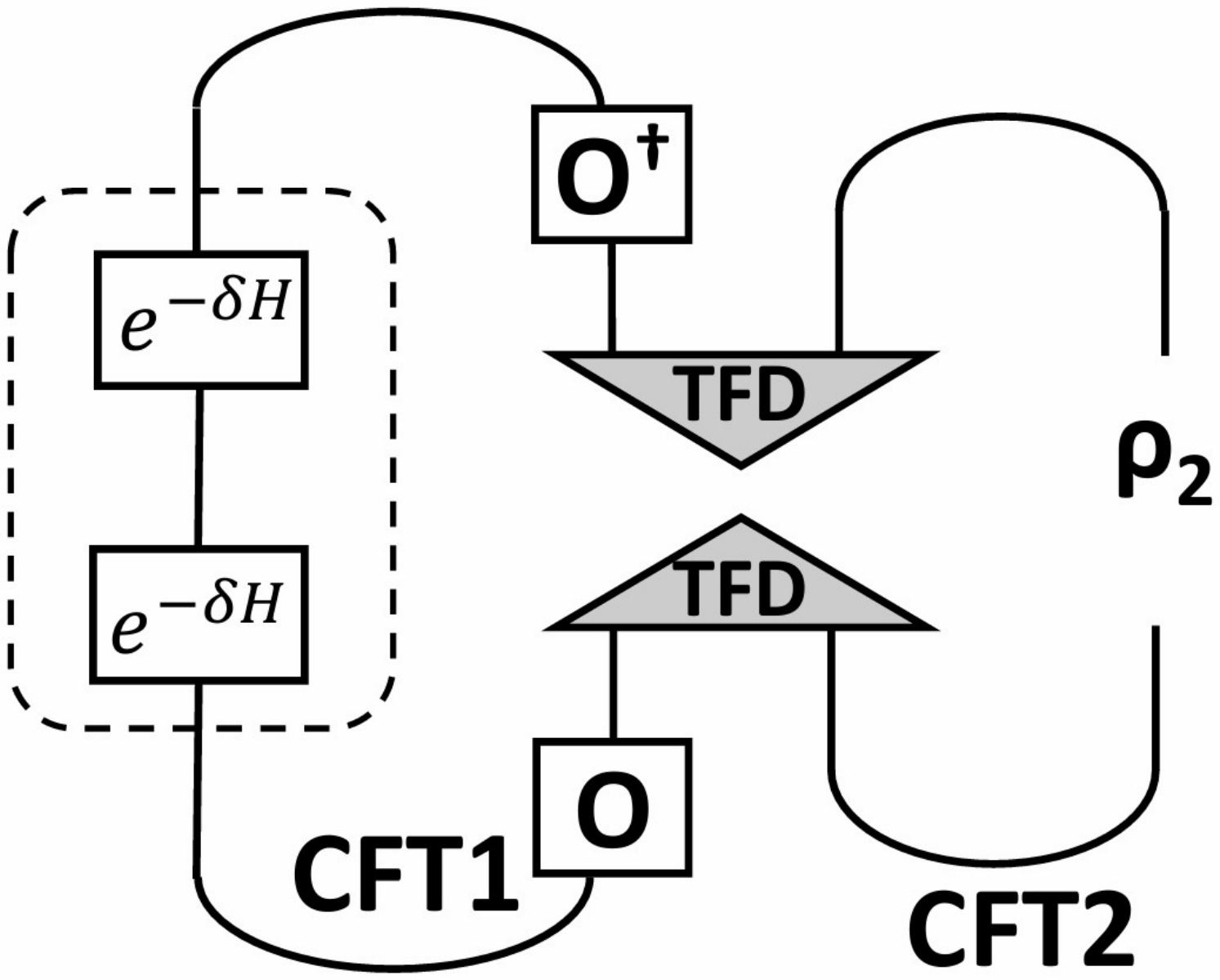}
    \caption{A sketch of the Euclidean path integral description of density matrix $\rho_2$ in CFT$_2$ (left) and a quantum circuit description of $\rho_2$ (right). If $\delta$ is large, the operation  surrounded by the dashed rectangle acts like a projection to the ground state and thus this becomes quantum teleportation of the state ${\cal O}|0\lb$ in CFT$_1$ to that in CFT$_2$.}
    \label{teleport}
\end{figure}

The main purpose of this paper is to provide an explicit and analytically tractable model of a pair of two-dimensional CFTs (called CFT$_1$ and CFT$_2$), where quantum entanglement and energy are transmitted from CFT$_1$ to CFT$_2$ as an interesting class of quantum operations.
A brief sketch of the construction of our model is as follows (refer to fig.\,\ref{teleport}). We consider a pair of identical CFTs on a two-dimensional spacetime and introduce quantum entanglement locally at time $t=t_M$ and location $x=0$. 
In the Euclidean path integral description, this localized entanglement is realized by gluing two small holes in the CFTs. The spatial size of the localized entanglement is described by parameter $s$, which serves as a UV regulator, and the effective inverse temperature $\beta$, which controls the conformal dimensions of local operators participating in the quantum entanglement. If the CFTs have a holographic dual, the gravitational counterpart to this model is an insertion of a wormhole, which can be thought of as a localized version of the eternal AdS black hole \cite{Maldacena:2001kr}. We will then introduce a local excitation ${\cal O}$ at time $t=0$ and location $x=x_P$ in CFT$_1$ and observe how this excitation is transmitted to CFT$_2$ under time evolution by computing the energy density and entanglement entropy. 
The spatial size of the localized excitation is described by parameter $\delta$. This local excitation is identical to the local operator quench, studied in \cite{NNT,Nozaki:2014hna,Nozaki:2014uaa,HNTW,Hat,CNT}.

For non-vanishing $\delta$, we can regard this as a version of quantum teleportation between CFT$_1$ and CFT$_2$, where the transmission becomes more successful as $\delta$ gets larger. A similar setup for a translationally invariant wormhole, i.e. the eternal AdS black hole, was considered in earlier papers \cite{Shenker:2013pqa,Shenker:2014cwa,Maldacena:2015waa,Caputa:2015waa} to study scrambling effects \cite{Hayden:2007cs,Sekino:2008he,Hosur:2015ylk,Swingle:2016var,Lashkari:2011yi} by computing the entanglement entropy, while no signal propagation between the two CFTs was considered as this wormhole is the eternal AdS black hole \cite{Maldacena:2001kr} and thus is not traversable. See also a recent paper \cite{Mao:2025scx} for a holographic CFT realization of the Hayden-Preskill model. In all of these arguments, the two CFTs are causally separated and we cannot any send signals from one to the other, which corresponds to the fact that the AdS wormhole is not traversable in the gravitational dual. 

However, the present paper argues, for the first time, that the transmission of signals from CFT$_1$ to CFT$_2$ does actually occur for non-zero $\delta$. This effect occurs because the Euclidean path integral is not a unitary evolution, which allows the signal to penetrate the black hole horizon. As we will see, this is directly related to the phenomenon opposite to the scrambling effect, where the correlation between two CFTs gets enhanced. We will also explain why these are possible by comparing it with quantum teleportation and also by closely following the propagation of quantum entanglement. Notice that this mechanism of our transmission between two CFTs is different from those of the traversable AdS black holes due to the double trance deformation \cite{Gao:2016bin,Maldacena:2018lmt} and the non-unitary Janus deformation \cite{Kawamoto:2025oko,Harper:2025lav}. 

This paper is organized as follows. In section \ref{Sec:Setup}, we explain the setup of our model and give a heuristic interpretation by relating it to quantum teleportation.
In section \ref{Sec:Energy}, we analyze the energy stress tensor to compute the energy transmission between the two CFTs.
In section \ref{Sec:EEMI}, we calculate the entanglement entropy and mutual information for various subsystems in our model with or without the local operator excitation and give qualitative interpretations.
In section \ref{Sec:Concl}, we summarize our findings and discuss future problems.
In appendix \ref{Append:Derivation}, we give a field-theoretic derivation of the entanglement entropy in a holographic theory when the subsystem is taken to be a double interval.

\section{A model of locally entangled CFTs}\label{Sec:Setup}
In this section, we present our model of the locally entangled CFTs via the Euclidean path integral and give an operational meaning in quantum information theory as a version of quantum teleportation. This model was briefly mentioned in appendix B of an earlier paper \cite{Numasawa:2016emc} and is closely related to the mixed-state local quench analyzed in \cite{Bhattacharyya:2019ifi,Doi:2025oma}.

\subsection{The localized thermofield double state}\label{Subsec:Desc}

To couple the two CFTs, we consider a time-evolving state analogous to the thermofield double (TFD) state inserted at $(t,x)=(t_M,0)$:
\begin{equation}\label{TFDstate}
    |\Psi(t-t_M)\lb_{12}=\sum_{i}\frac{e^{-\beta (\Delta_i-\frac{c}{12})}}{\s{Z(2\beta)}}
    |{\cal O}_i(-t+t_M)\lb_1 |{\cal O}_i(-t+t_M)\lb_2.
\end{equation}
where $2\beta$ is the inverse temperature, $Z(2\beta)=\sum_{i}e^{-2\beta\left(\Delta_i-\frac{c}{12}\right)}$ is the partition function, and $O_i$ denotes each of all operators in the theory with conformal dimension $\Delta_i$. The states $|O_i(t)\lb$ are UV-regularized so that
\begin{equation}
    |\mathcal{O}_i(t)\lb={\cal N}_i e^{-sH}\mathcal{O}_i(t)|0\lb,
\end{equation}
where $H$ is the Hamiltonian of the CFT, $s$ serves as the regulator, and $\mathcal{O}_i(t)=e^{itH}\mathcal{O}_ie^{-itH}$ as usual. The normalization ${\cal N}_i$ is chosen such that 
$\la \mathcal{O}_i(t) |\mathcal{O}_j(t)\lb=\delta_{i,j}$. The sum labeled by $i$ in (\ref{TFDstate}) runs over all states, including descendants of primary states. We shall call this state the ``localized TFD state''.

\begin{figure}[h]
    \centering
    \includegraphics[width=0.6\linewidth]{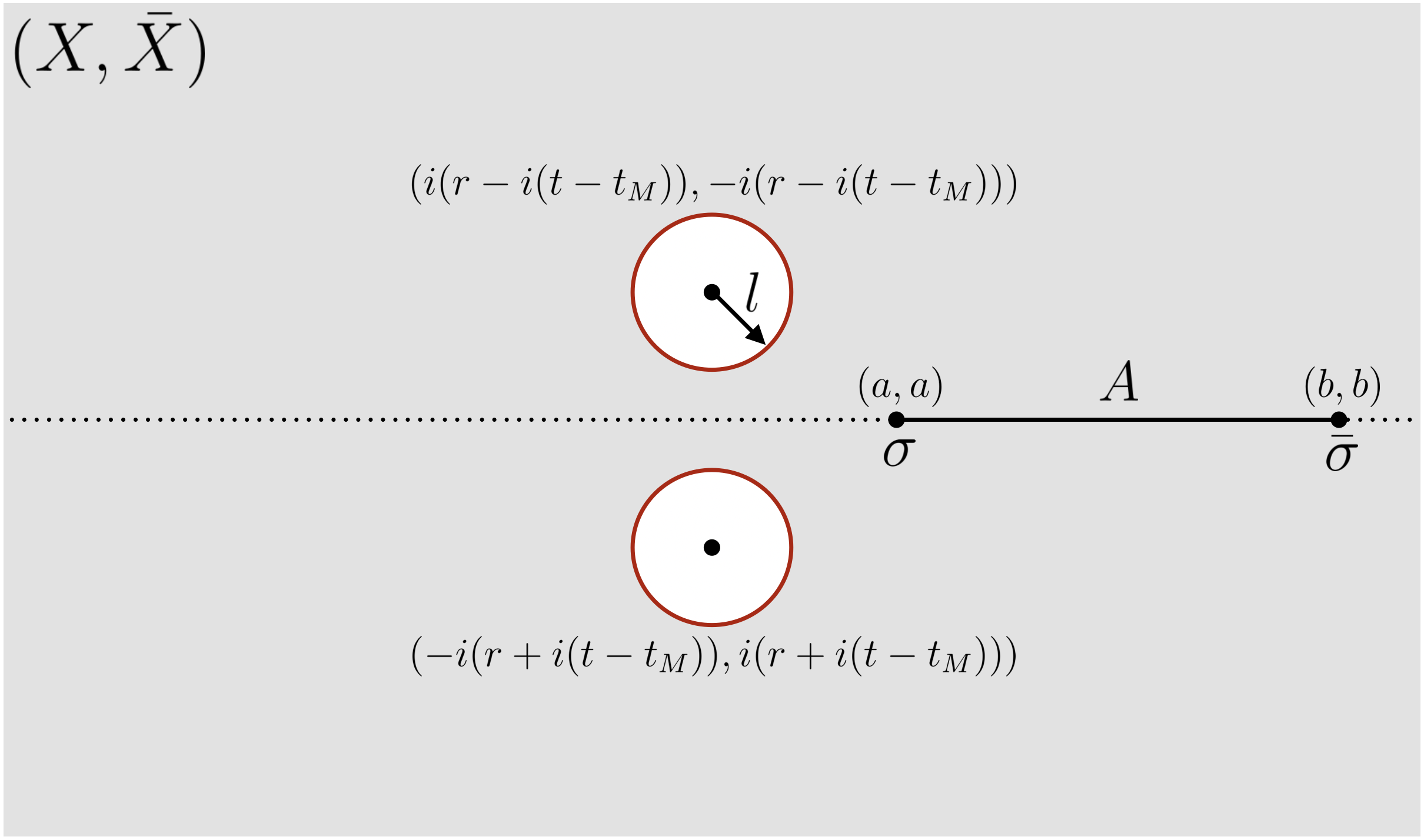}
    \caption{Euclidean path integral over a plane with two holes inserted. We prepare two copies of the same sheet and identify the circles.}
    \label{two-hole-plane}
\end{figure}

In the Euclidean path integral formalism, such a state can be prepared by inserting a hole in two half-planes (one per CFT) and gluing them together \cite{Numasawa:2016emc,Bhattacharyya:2019ifi,Doi:2025oma}. To be specific, the path integral is performed over two planes that are glued together at locations 
\begin{align} 
    &X_1=i(r-i(t-t_M))\,, &\bar{X}_1&=-i(r-i(t-t_M)), \\
    &X_2=-i(r+i(t-t_M))\,, &\bar{X}_2&=i(r+i(t-t_M)) ,
    \label{locap}
\end{align}
where $r=s/\tanh(\beta/2)$, forming a geometry that is conformally equivalent to a torus, see fig.\,\ref{two-hole-plane}. The radius of the holes is to be set at $l=s/\sinh(\beta/2)$. As noted in \cite{Doi:2025oma}, there is a one-to-one correspondence between the parameters $(r,l)$ and $(\beta,s)$. Either set of parameters can be used to fully define the state (\ref{TFDstate}), and since we would like to fix $\beta$ in this paper, we shall adopt the latter hereafter.
If the CFT has a holographic dual, this setup can be interpreted as a localized black hole or wormhole that is falling in AdS space \cite{Maldacena:2001kr}.
Let us also comment that this setup is in some sense a purification of the setup of the mixed-state local quench in \cite{Doi:2025oma}.

\begin{figure}[h]
    \centering
    \includegraphics[width=0.4\linewidth]{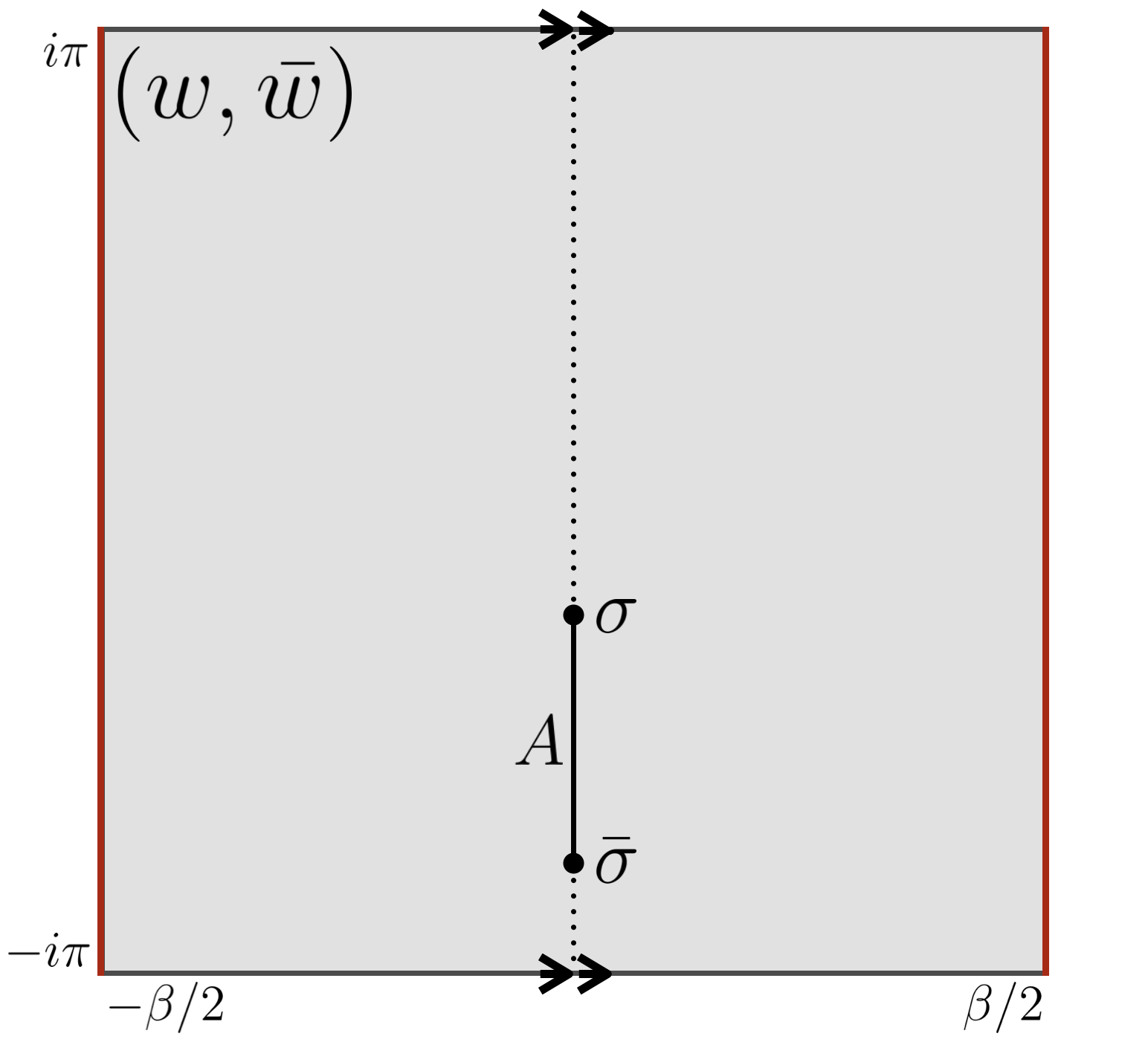}
    \caption{The $(w,\bar{w})$ strip. The left and right edges of this strip correspond to the circles in the original $(X,\bar{X})$ geometry. The wormhole, i.e. the torus, is obtained by taking two copies of the strip and identifying the left and right edges of one of the copies to the corresponding edges of the other.}
    \label{strip}
\end{figure}

In general, computation of correlation functions on a torus requires details of the CFT of interest. We shall examine free CFTs as well as holographic CFTs in this paper. However, regardless of the theory in question, it would be beneficial to work with a nicer set of coordinates instead of $(X,\bar{X})$. First, map the punctured planes to strips via the conformal transformation
\begin{align} \label{eq:X2w}
    w=\log\left(-\frac{X-t+t_M+is}{X-t+t_M-is}\right)\,, \quad
    \bar{w}=\log\left(-\frac{\bar{X}+t-t_M-is}{\bar{X}+t-t_M+is}\right)
\end{align}
and denote the coordinates in CFT\textsubscript{1} and in CFT\textsubscript{2} by $w_1$ and $w_2$ respectively. The real part of $w$ ranges from $-\beta/2$ to $\beta/2$ and the imaginary part ranges from $-i\pi$ to $i\pi$ as can be seen in fig.\,\ref{strip}.

\begin{figure}
    \centering
    \includegraphics[width=0.4\linewidth]{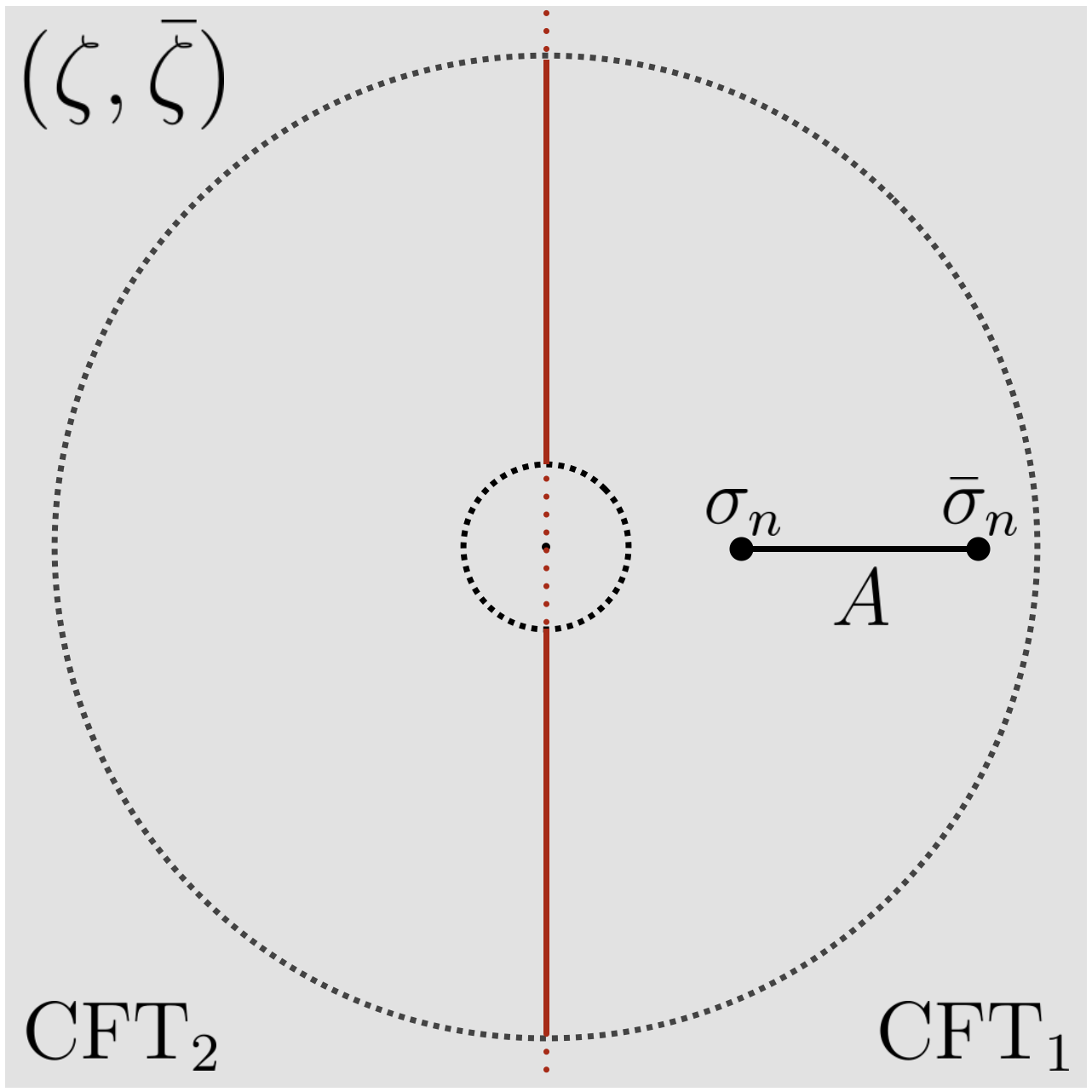}
    \caption{Annulus formed by decompactifying the torus. CFT\textsubscript{1} is mapped to the right half and CFT\textsubscript{2} is mapped to the left half.}
    \label{annulus}
\end{figure}

For holographic CFTs, since there is a sharp phase transition in the torus partition function at $\beta=\pi$, which corresponds to the Hawking-Page transition in its gravitational dual, we can pick a phase of our liking and approximate the torus as an infinite cylinder. This involves a decompactification of either the temporal or spatial direction of the torus. In this paper, we shall focus on the high temperature phase $\beta<\pi$ and decompactify the spatial direction by the conformal map
\begin{equation} \label{eq:w2zeta}
    \zeta=\left\{
    \begin{aligned}
        &e^{\frac{\pi i}{\beta}w} \,, &w\in\text{CFT}_1\\
        &-e^{-\frac{\pi i}{\beta}w}\,,&w\in\text{CFT}_2
    \end{aligned}\right.
\end{equation}
which results in an annulus; see fig.\,\ref{annulus}. The spatial direction now extends radially in the annulus and its decompactification gives an entire Euclidean plane, on which we know how to compute correlation functions. Finally, correlation functions obtained on the $(\zeta,\bar{\zeta})$ plane can be conformally transformed back to those in terms of the $(X,\bar{X})$-coordinates in CFT\textsubscript{1} and CFT\textsubscript{2} respectively. Note that on the time slice where the two CFTs are glued together, the spatial orientations of the CFTs are opposite to each other, as $w_1=\beta/2+ix \sim w_2=\beta/2-ix$ and $w_1=-\beta/2+ix \sim w_2=-\beta/2-ix$ for all $x \in [-\pi,\pi]$.

For later convenience, let us also briefly consider the structure of quantum entanglement in the localized TFD state (\ref{TFDstate}). To understand this, it is important that after the time evolution of the localized TFD state inserted at time $t_M$, it entangles the following four points at time $t$:
\ba
&& P_1: x=t-t_M\ \  \mbox{in CFT}_1,\no
&& Q_1: x=-(t-t_M)\ \  \mbox{in CFT}_1,\no
&& Q_2: x=t-t_M\ \  \mbox{in CFT}_2,\no
&& P_2: x=-(t-t_M)\ \  \mbox{in CFT}_2.  \label{fourp}
\ea
This argument can be understood if we remember that the localized TFD state looks like (\ref{TFDstate}), where we can view the locally excited states $\ket{{\cal O}_i(t)}_{(1)}$ in CFT$_1$ and $\ket{{\cal O}_i(t)}_{(2)}$ in CFT$_2$ as
\ba
&& \ket{{\cal O}_i(t)}_{(1)}=\sum_{\ap}{\cal O}^{(\ap)}_{iL}(t){\cal O}^{(\ap)}_{iR}(t)|0\lb_{(1)},\no
&& \ket{{\cal O}_i(t)}_{(2)}=\sum_{\ap}{\cal O}^{(\ap)}_{iL}(t){\cal O}^{(\ap)}_{iR}(t)|0\lb_{(2)}.  \label{localent}
\ea
Here ${\cal O}^{(\ap)}_{iL}$ and ${\cal O}^{(\ap)}_{iR}$ denote the left and right-moving parts of the local excitation. Under time evolution of the two CFTs, the left and right-moving modes propagate at the speed of light in opposite directions of $x$.
Thus, bipartite quantum entanglement is generated for each pair of the four points $P_1,Q_1,P_2$ and $Q_2$ in (\ref{fourp}).




\begin{figure}
    \centering
    \includegraphics[width=0.7\linewidth]{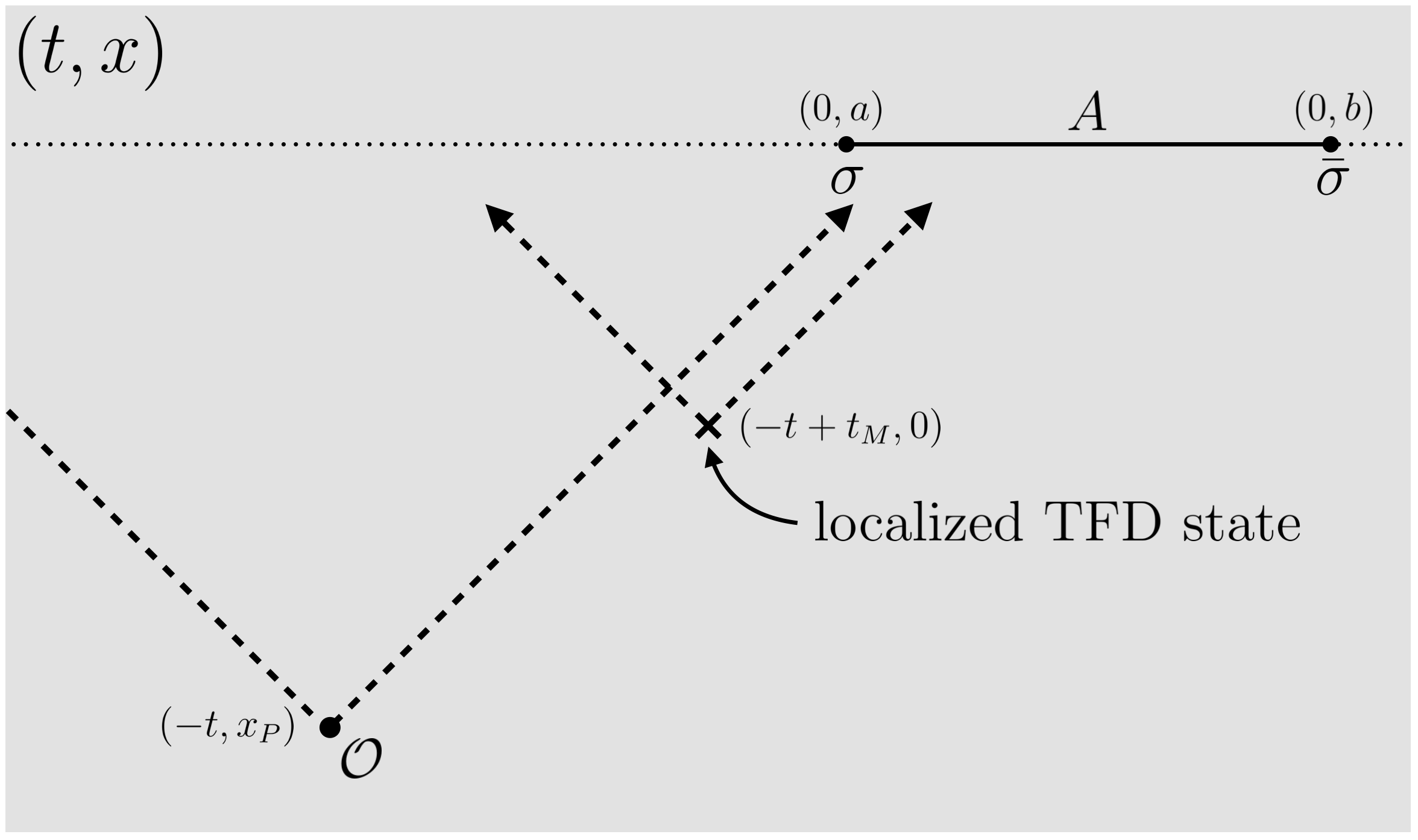}
    \caption{The Lorentzian setup of the local operator quench, represented by the operator insertion $\mathcal{O}$ at $(t,x)=(0,x_P)$. The background is defined by the localized TFD state inserted at $(t,x)=(t_M,0)$.}
    \label{fig:lorentziansetup}
\end{figure}

To model an interesting class of quantum operations, we further insert a local operator ${\cal O}$ at time $t=0$ and location $x=x_P$ in CFT$_1$ as depicted in fig.\,\ref{fig:lorentziansetup}. Since the local operator is singular and has infinite energy, we regulate it by the damping operation $e^{-\delta H}$. The resulting state is expressed as
\begin{equation}\label{opstate}
    |\Psi^{\text{quench}}(t)\lb_{12}={\cal N} \cdot e^{-\delta H^{(1)}} {\cal O}(x_P,-t) e^{\delta H^{(1)}}|\Psi(t-t_M)\lb_{12}
\end{equation}
where the normalization ${\cal N}$ is chosen so that $_{12}\langle\Psi^{\text{quench}}(t)|\Psi^{\text{quench}}(t)\rangle_{12}=1$. We can identify this excited state as a local operator quench in the localized TFD state. Throughout this paper, we shall assume $0<\delta <s$. In principle, we can also consider $s<\delta$, but as pointed out in \cite{Doi:2025oma}, computational difficulties arise due to the ambiguity in logarithmic branches and the lack of a clear prescription to select the correct one.

\begin{figure}
    \centering
    \includegraphics[width=0.7\linewidth]{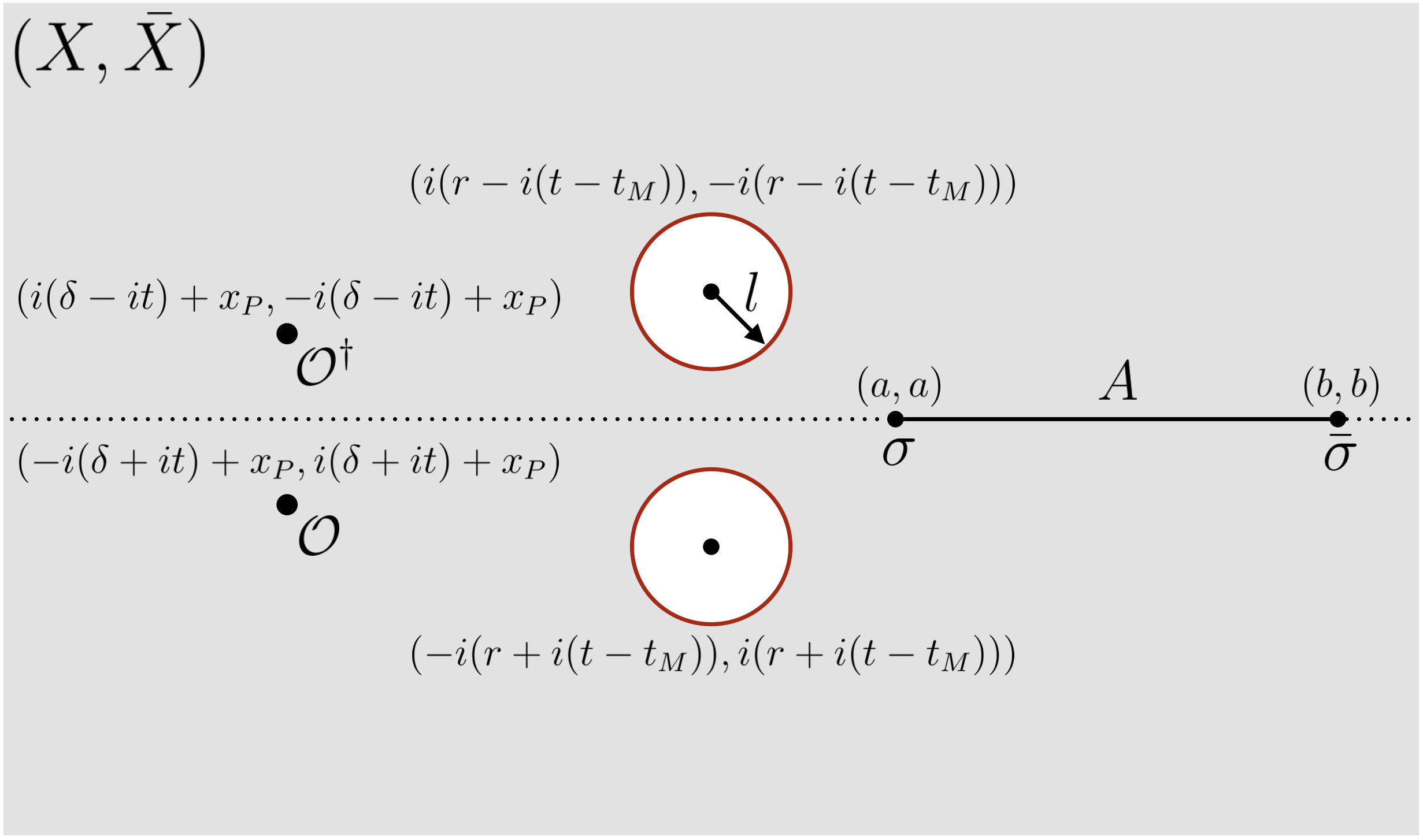}
    \caption{Euclidean representation of the local operator quench in the localized TFD state.}
    \label{fig:localquench}
\end{figure}

Once again in the Euclidean path integral formalism, the local quench can be realized by a pair of operator insertions at
\begin{align}
    X_1 &= x_P+i(\delta-it), & \bar{X}_1 &= x_P-i(\delta-it), \\
    X_2 &= x_P-i(\delta+it), & \bar{X}_2 &= x_P+i(\delta+it),
\end{align}
in the two-hole background geometry; this is depicted in fig.\,\ref{fig:localquench}.

We will shortly discuss the operational meaning of the state $|\Psi^{\text{quench}}(t)\lb_{12}$ as a form of quantum teleportation (refer to section \ref{Subsec:QT}) and later explore its properties quantitatively (refer to sections \ref{Sec:Energy} and \ref{Sec:EEMI}).

\subsection{Connection to quantum teleportation}\label{Subsec:QT}

Returning to state (\ref{opstate}), the reduced density matrix in CFT$_{2}$ is given by
\begin{equation}
    \rho_2 = {\cal N}^2\cdot \text{Tr}_1\left[ \text{ } e^{-\delta H^{(1)}} {\cal O}(x_P,-t) e^{\delta H^{(1)}}|\Psi(t-t_M)\lb_{12}\la\Psi(t-t_M)|_{12}
    e^{\delta H^{(1)}}{\cal O}^\dagger(x_P,-t)
    e^{-\delta H^{(1)}}\right].
\end{equation}
The Euclidean path integral description of this density matrix is sketched in the left panel of fig.\,\ref{teleport}.

Plugging in the expression (\ref{TFDstate}) for $|\Psi\lb_{12}$ into the density matrix gives
\ba
    \rho_2 &=& \frac{{\cal N}^2}{Z(2\beta)}\sum_{ij} {\cal N}_i^2 {\cal N}_j^2 e^{-\beta\left(\Delta_i+\Delta_j-\frac{c}{6}\right)}
\la 0|{\cal O}^\dagger_i(x_1,t_M-t)e^{-(s-\delta)H^{(1)}} 
{\cal O}^\dagger(x_P,-t)e^{-2\delta H^{(1)}}{\cal O}(x_P,-t)\no
&& \ \ \ \ \times e^{-(s-\delta)H^{(1)}}{\cal O}_j(x_1,t_M-t)|0\lb\otimes  \left[
 e^{-sH^{(2)}}{\cal O}_j(x_2,t_M-t)|0\lb \la 0| {\cal O}^\dagger_i(x_2,t_M-t)e^{-sH^{(2)}}\right].\no
\ea
Here $x_1$ and $x_2$ are the points in CFT$_1$ and CFT$_2$ that are entangled with each other in the localized TFD state. In this paper we simply set $x_1=x_2=0$ as in (\ref{locap}).

We see that when $\delta \ll s$, the four function decouples into $\bra{{\cal{O}}_i} {\cal O}^\dagger {\cal O} \ket{{\cal O}_j} \approx \langle {\cal O}^\dagger {\cal O} \rangle \delta_{ij}$. In this case, the density matrix is essentially the same as the state without the local operator excitation. 

On the other hand, if one looks at the case where we have  $\delta \simeq s$ and  $|x_P+t_M|\ll \delta$, then the channels ${\cal O}_j\times{\cal O}^\dagger$ and ${\cal O}\times{\cal O}_i$ dominate. Therefore the summations over $i$ and $j$ becomes localized when $\mathcal{O}_i=\mathcal{O}_j=\mathcal{O}$. In this case, the density matrix is approximately
\begin{equation}
    \rho_2 \simeq |{\cal{O}}\lb \la {\cal{O}}|.
\end{equation}
Thus the local excitation ${\cal O}$ inserted in CFT$_1$ is now teleported to CFT$_2$ successfully.

This operation is sketched in the right panel of fig.\,\ref{teleport} as a quantum circuit.
When $\delta \gg s-\delta$, the operation $e^{-\delta H}$ looks like a projection to the ground state $|0\lb$. In this limit, the protocol is identical to quantum teleportation. The input state $O(x_P)|0\lb_{(1)}$ in CFT$_1$ is teleported to $O(x_P)|0\lb_{(2)}$ in CFT$_2$ by the projection to the Bell state, which is the localized TFD state $|\Psi\lb_{12}$ in our model.

\section{Energy transmission}\label{Sec:Energy}
As the first step to the analysis of our model of the locally entangled system with a local operator excitation, here we calculate the energy density in CFT$_1$ and CFT$_2$ to see how the local excitation propagate from CFT$_1$ to CFT$_2$. We will analyze both holographic CFTs and the free scalar below.

\subsection{Holographic CFT}\label{Subsec:EnergyHol}
In this section, we will evaluate the energy density distribution in the two conformal field theories in the case where both of them are the same holographic conformal field theory.
The calculation proceeds in a similar fashion to \cite{CNT,Doi:2025oma}.

One first maps the two planes onto a torus.
In the decompactified limit, one can then approximate the torus as a cylinder.
The energy density is computed as the sum of the holomorphic and antiholomorphic parts of the stress-energy tensor. For the holomorphic part, we find that at $(X,\bar{X})=(x,x)$,
\begin{align}
    & \frac{\langle T(x,x) \mathcal{O}^\dagger(i(\delta-it)+x_P,-i(\delta-it)+x_P) \mathcal{O}(-i(\delta+it)+x_P,i(\delta+it)+x_P)\rangle}{\langle \mathcal{O}^\dagger(i(\delta-it)+x_P,-i(\delta-it)+x_P) \mathcal{O}(-i(\delta+it)+x_P,i(\delta+it)+x_P)\rangle} \nn\\
    &= \frac{4\pi^2 hs^2}{\beta^2 ((x-t+t_M)^2 + s^2)^2} \frac{\zeta(x)^2(\zeta_{\mathcal{O}^\dagger}-\zeta_{\mathcal{O}})^2}{(\zeta(x)-\zeta_{\mathcal{O}^\dagger})^2(\zeta(x)-\zeta_{\mathcal{O}})^2},
\end{align}
excluding the term associated with the Schwarzian derivative arising from the transformation between $(X,\bar{X})$ and $(\zeta,\bar{\zeta})$.
Note that the Schwarzian derivative contribution corresponds to the energy density of the localized TFD state. 
$h$ is the scaling dimension of the operators $\mathcal{O}, \mathcal{O}^\dagger$.
Likewise, the antiholomorphic part of the stress-energy tensor can be computed as follows:
\begin{align}
    & \frac{\langle \bar{T}(x,x) \mathcal{O}^\dagger(i(\delta-it)+x_P,-i(\delta-it)+x_P) \mathcal{O}(-i(\delta+it)+x_P,i(\delta+it)+x_P)\rangle}{\langle \mathcal{O}^\dagger(i(\delta-it)+x_P,-i(\delta-it)+x_P) \mathcal{O}(-i(\delta+it)+x_P,i(\delta+it)+x_P)\rangle} \nn\\
    &= \frac{4\pi^2 hs^2}{\beta^2 ((x+t-t_M)^2 + s^2)^2} \frac{\bar{\zeta}(x)^2(\bar{\zeta}_{\mathcal{O}^\dagger}-\bar{\zeta}_{\mathcal{O}})^2}{(\bar{\zeta}(x)-\bar{\zeta}_{\mathcal{O}^\dagger})^2(\bar{\zeta}(x)-\bar{\zeta}_{\mathcal{O}})^2}.
\end{align}
Note that the specific form of the maps $\zeta(X), \bar{\zeta}(\bar{X})$ are given by equations \eqref{eq:X2w} and \eqref{eq:w2zeta} depending on whether $(X, \bar{X})$ is in CFT$_1$ or CFT$_2$.
As a remark, since we have employed the approximation for which the torus becomes a cylinder, the functional form of the energy density is essentially universal, depending on the local operator only through the scaling dimension $h$ appearing as an overall prefactor.
Using these formulas, one can then go on to numerically evaluate the energy density distribution and its time evolution.
\begin{figure}[t]
    \centering
    \begin{minipage}{0.45\linewidth}
        \centering 
        \includegraphics[width = \linewidth]{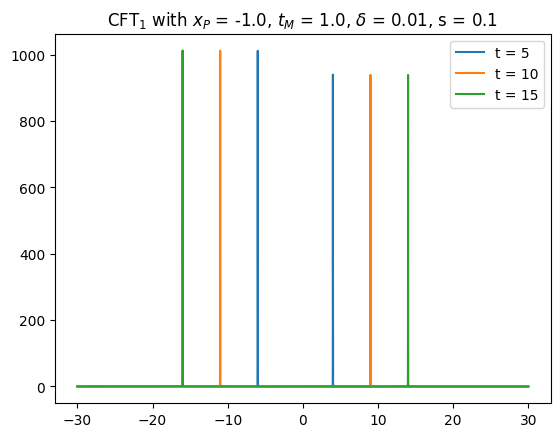}
        \caption{Time evolution of energy density in CFT$_1$. We subtracted the contribution from the localized TFD state.}
        \label{sec3:cft1_te}
    \end{minipage}
    \hspace{1cm}
    \begin{minipage}{0.45\linewidth}
        \centering 
        \includegraphics[width = \linewidth]{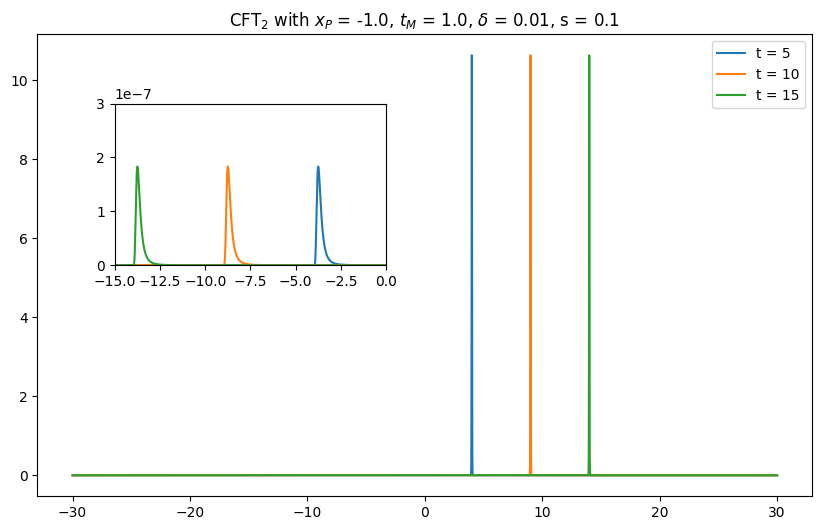}
        \caption{Time evolution of energy density in CFT$_2$. We subtracted the contribution from the localized TFD state. In addition to the right-moving peak, there is a very small left-moving peak hidden in the graph.}
        \label{sec3:cft2_te}
    \end{minipage}
\end{figure}

\begin{figure}
    \centering
    \includegraphics[width=8cm]{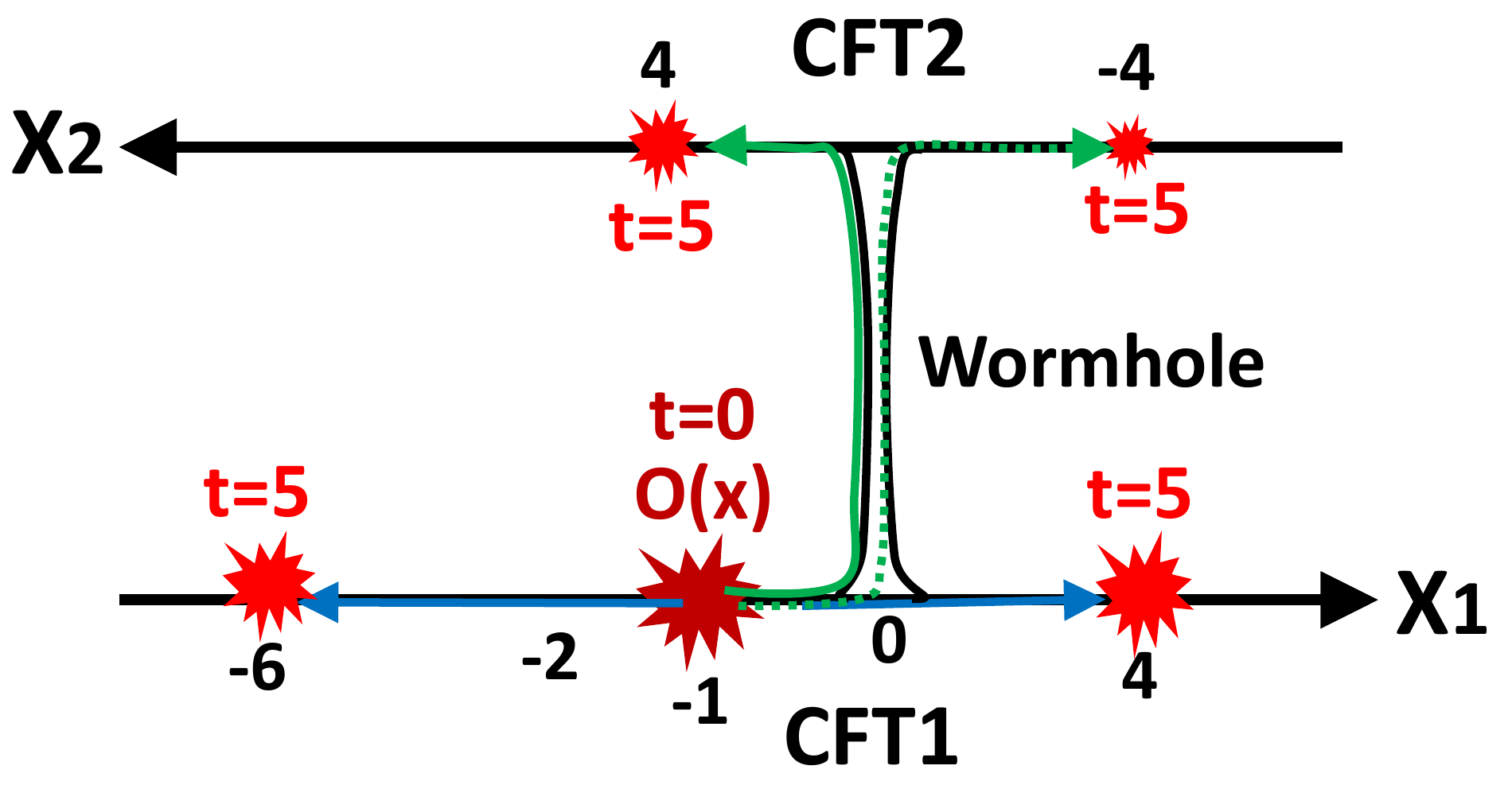}
    \caption{A sketch of propagation of the local operator excitation from $t=0$ to $t=5$.}
    \label{fig:propa}
\end{figure}

The time evolution of energy density for both CFTs is given in fig.\,\ref{sec3:cft1_te} and fig.\,\ref{sec3:cft2_te}.
From here until the end of this section, we set $\beta = 1$.
The plot shows the case where the local quench is lightlike-separated from the localized TFD state.
This is the case for which transmission is expected to be maximal because the local excitation hits the center of the wormhole. On CFT$_1$, one can see that there are two peaks moving away from each other as time evolves.
This corresponds to the two modes excited by the local operator.
On CFT$_2$, though only a single peak is visible there is also a tiny peak hidden on the left.
This means that a small portion of the energy in both of the two modes in CFT$_1$ get transported to the other CFT.
Note that the transmission is not strong as the amplitude on the second CFT is much smaller than the first one, which is expected given the parameters and the discussion from the previous section. This evolution is depicted in fig.\,\ref{fig:propa}.

From fig.\,\ref{sec3:cft1_te} and fig.\,\ref{sec3:cft2_te}, one can also see that the amplitude and shape of the peaks do not change as a function of time.
This is consistent with unitary evolution after the quench.
One can check to see that the total energy of the left-moving peak in CFT$_1$ is not equal to the sum of the right-moving peak in CFT$_1$ and the peaks transmitted to CFT$_2$.
This can be understood as some of the energy leaking out of the system during the quench.

\begin{figure}[h]
    \centering
    \begin{minipage}{0.45\linewidth}
        \includegraphics[width = \linewidth]{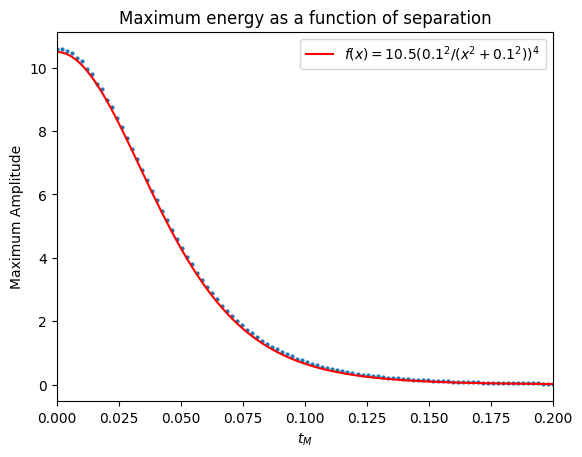}
    \end{minipage}
    \hspace{1cm}
    \begin{minipage}{0.45\linewidth}
        \begin{tikzpicture}[decoration = snake]
        \draw[-{Stealth[length=2.5mm]}] (-3, 0) -- (3, 0);

        \filldraw [black] (0, 0) circle (1pt) node[above] {$x = 0$};
        \filldraw [black] (0, -2.1) circle (2.5pt) node[below] {local operator quench $O$};
        \draw[<->, dotted, very thick] (0, -2) -- (0, -0.1);
        \node[right] at (0.1, -0.75) {$t_M$};
        \draw[->, decorate, red, ultra thin] (+0.2, -1.8) -- (1, -1);
        \draw[->, decorate, red, ultra thin] (-0.2, -1.8) -- (-1, -1);
        \end{tikzpicture}
    \end{minipage}
    \caption{Left: Maximum energy density on the second CFT; Right: Setup being used to calculate energy transfer: the local operator is kept at $x_P = 0$ while the time separation between wormhole and local quench $t_M$ is changed.}
    \label{sec3:enetransfer}
\end{figure}
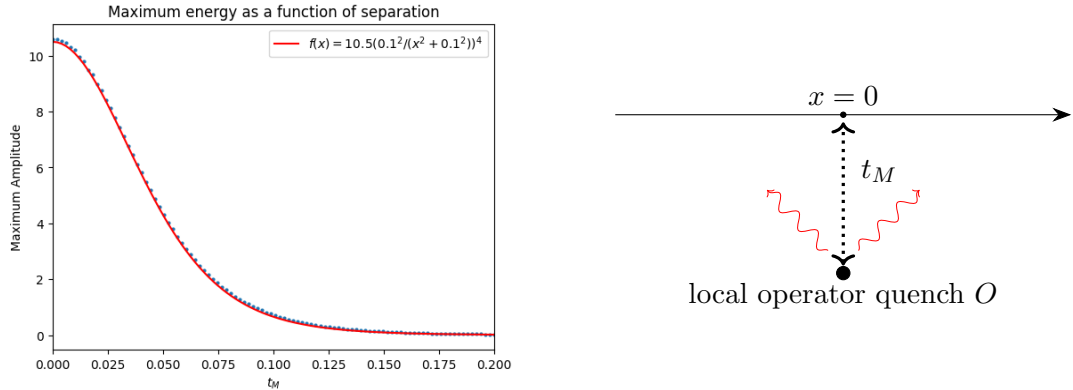

As one moves away from lightlike separation, the amount of energy transferred to the second CFT decreases. 
Consider the setup shown on the right of fig.\,\ref{sec3:enetransfer}.
For this setup, one will observe that there are two peaks in CFT$_2$ moving away from the origin.
When $t_M > 0$, the separation is not lightlike and the energy transfer is not maximal.
As the wormhole has a finite size given by $s$, some of the energy can still pass through the second CFT.
From the plot, it is possible to see that the decay rate is controlled by $s$ --- the size of the wormhole (note that for this calculation, we have set $\delta = 0.01, s = 0.1$).
The blue dots show the calculation results and the red line is a function that fits the results well.

\begin{figure}
    \centering
    \includegraphics[width=0.5\linewidth]{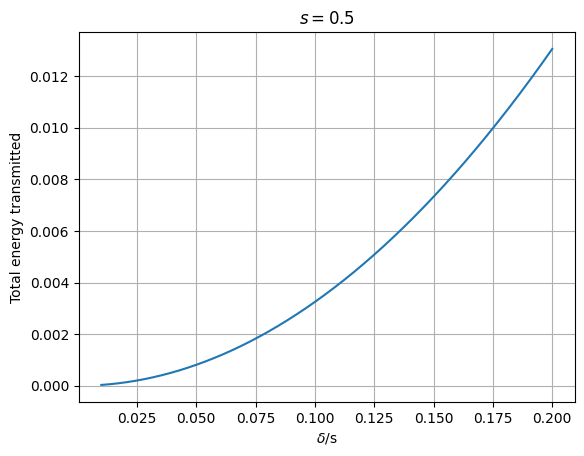}
    \caption{Energy transfer as a function of $\delta/s$}
    \label{sec3:relativesize}
\end{figure}

It is also interesting to look at how the energy transfer varies as the ratio $\delta/s$ is changed.
Fig.\,\ref{sec3:relativesize} shows the total amount of energy transferred to the second CFT as $\delta$ is varied while keeping $s = 0.1$.
The other parameters are also kept fixed and we are working in the null separation case, which is similar to the situation shown in fig.\,\ref{sec3:cft1_te} and fig.\,\ref{sec3:cft2_te}.
One can see that as $\delta$ is increased, the amount of energy going through the wormhole also increases, consistent with the expectation in the previous section.
Note that we are keeping the ratio small as the approximation that we are using breaks down when $\delta/s=O(1)$.

\subsection{Free CFT}\label{Subsec:EnergyFree}
In this section, we will evaluate the energy distribution for the free bosonic CFT.
Since the two-point functions on the torus for certain operators are known, one can apply the Ward identity to get the energy distribution without resorting to any approximations.
An exact calculation on the torus means that the three-point function will depend on the operator content of the specific theory and one loses the universal result obtained in the previous section.

Before going into specific models, let us first outline the setup.
First, each CFT is mapped to a cylinder (with coordinate $w$) through the given map.
Then, one glues the two cylinders into a torus as follows.
Denote the coordinates by $z, \bar{z}$, one has $ -\frac{1}{2}\leq\Re z, \Re \bar{z} \leq \frac{1}{2}, -\frac{\beta}{2\pi} \leq \Im z, \Im \bar{z} \leq \frac{\beta}{2\pi}$ (that is, $\tau = \frac{i\beta}{\pi}$).
Let $(w_1, \bar{w}_1), (w_2, \bar{w}_2)$ be the coordinates on the cylinder for CFT$_1$ and CFT$_2$ respectively, we then have
\begin{align}
    z &= \begin{cases}
        \frac{i}{2\pi}(w_1 - \frac{\beta}{2}) & \text{CFT}_1 \\
        \frac{i}{2\pi}(-w_2 + \frac{\beta}{2}) & \text{CFT}_2
    \end{cases} \\
    \bar{z} &= \begin{cases}
        -\frac{i}{2\pi}(\bar{w}_1 - \frac{\beta}{2}) & \text{CFT}_1 \\
        -\frac{i}{2\pi}(-\bar{w}_2 + \frac{\beta}{2}) & \text{CFT}_2
    \end{cases} 
\end{align}
This mapping is done so that one obtains a parametrization of the torus that agrees with the one used in \cite{DiFrancesco:1997nk}.

Let $\mathcal{O}, \mathcal{O}^\dagger$ be primary operators of scaling dimension $h$. 
The Ward identity (for the holomorphic part) on the torus gives (see, for example, chapter 10 of \cite{DiFrancesco:1997nk})
\begin{equation}\label{eq.Torus_Ward_Id}
    \begin{split}
    \frac{\langle T(z)\mathcal{O}(z_1)\mathcal{O}^\dagger(z_2) \rangle}{\langle \mathcal{O}(z_1) \mathcal{O}^\dagger(z_2) \rangle} =& \langle T \rangle + 4h\eta_1 + h(\wp(z-z_1) + \wp(z-z_2)) \\
    &+(\zeta(z-z_1) - \zeta(z-z_2) + 2\eta_1 z_{12}) \partial_{z_{12}} \log \langle \mathcal{O}(z_1) \mathcal{O}^\dagger(z_2) \rangle \\
    &+ 2\pi i \partial_\tau \log \langle \mathcal{O}(z_1) \mathcal{O}^\dagger(z_2) \rangle
    \end{split}
\end{equation}
where 
\begin{equation}
    \zeta(z) = \frac{\partial_z \theta_1(z|\tau)}{\theta_1(z|\tau)} + 2\eta_1 z, \quad \quad
    \wp(z) = -\partial_z \zeta(z), \quad\quad
    \eta_1 = - \frac{1}{6} \frac{\partial^3_z \theta_1(0|\tau)}{\partial_z \theta_1(0|\tau)}
\end{equation}
with 
\begin{equation}
    \theta_1(z|\tau) = -i \sum_{r \in \mathbb{Z} + \frac{1}{2}} (-1)^{r-\frac{1}{2}} y^r q^{\frac{r^2}{2}}.
\end{equation}
Here, $y = e^{2\pi i z}, \ q = e^{2\pi i \tau}$. 
One has a similar expression for the antiholomorphic part.

This expression for the correlator on the $(z, \bar{z})$ torus can then be related to the holomorphic energy density on the $(X, \bar{X})$ plane through the relation
\begin{equation}
    \frac{\langle T(x) \mathcal{O}(X_1) \mathcal{O}^\dagger(X_2) \rangle_{X,\bar{X}}}{\langle \mathcal{O}(X_1) \mathcal{O}^\dagger(X_2) \rangle_{X, \bar{X}}} = \bigg(\frac{1}{\pi}\frac{s}{(x-t+t_M)^2+s^2}\bigg)^2 \frac{\langle T(z)\mathcal{O}(z_1)\mathcal{O}^\dagger(z_2) \rangle_{z, \bar{z}}}{\langle \mathcal{O}(z_1) \mathcal{O}^\dagger(z_2) \rangle_{z, \bar{z}}}.
\end{equation}
A similar expression also holds for the antiholomorphic part and the total energy density is simply the sum of both the holomorphic and antiholomorphic parts.
Here we have omitted the Schwarzian derivative term as well as we want to focus on the contribution coming from the local operator.

Now we will specialize to the case of the free noncompact boson.
We shall take the local operator to be $\mathcal{O} = e^{ip\phi}$, with $\phi$ being the fundamental scalar field.
The scaling dimension for such an operator is given by $h = \frac{p^2}{2}$.
The two-point function for $\mathcal{O}$ on the torus is given by
\begin{equation}
    \langle \mathcal{O}(z) \mathcal{O}^\dagger(0) \rangle = \bigg(\frac{\partial_z \theta_1(0|\tau)}{\theta_1(z|\tau)}\bigg)^{2h} \bigg(\frac{\partial_{\bar{z}} \theta_1(0|-\bar{\tau})}{\theta_1(\bar{z}|-\bar{\tau})}\bigg)^{2h} \exp{\frac{2\pi p^2 (\Im z)^2}{\Im \tau}}.
\end{equation}
Remember that we are taking $\tau = -\bar{\tau} = \frac{i\beta}{\pi}$.
Plugging this into the Ward identity, one obtains the following expression for the three-point correlator:
\begin{equation}
    \begin{split}
    \frac{\langle T(z)\mathcal{O}(z_1)\mathcal{O}^\dagger(z_2) \rangle}{\langle \mathcal{O}(z_1) \mathcal{O}^\dagger(z_2) \rangle} =& \langle T \rangle + 4h\eta_1 + h\bigg(\wp(z-z_1) + \wp(z-z_2)\bigg) \\
    &-(\zeta(z-z_1) - \zeta(z-z_2) + 2\eta_1 z_{12}) \bigg(2h\frac{\partial_z \theta_1(z_{12}|\tau)}{\theta_1(z_{12}|\tau)}+2\pi i p^2 \frac{\Im z_{12}}{\Im\tau}\bigg)\\
    &h \bigg(\frac{\partial_z^3 \theta_1(0|\tau)}{\theta_1(0|\tau)} - \frac{\partial_z^2 \theta_1(z_{12}|\tau)}{\theta_1(z_{12}|\tau)}\bigg) -2\pi p^2\left(\frac{\Im z_{12}}{\Im\tau}\right)^2.
    \end{split}
\end{equation}
\begin{figure}[t]
    \centering
    \begin{minipage}{0.45\linewidth}
        \includegraphics[width = \linewidth]{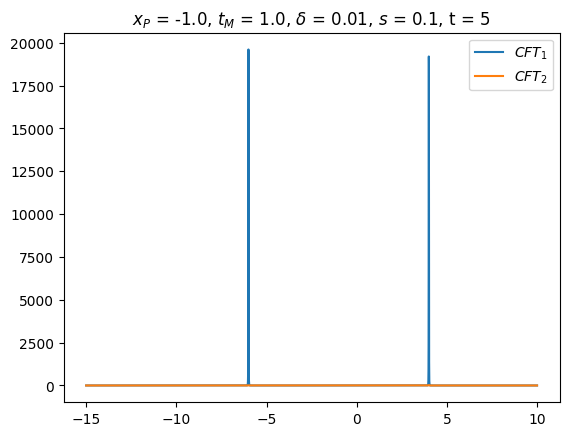}
    \end{minipage}
    \hspace{1cm}
    \begin{minipage}{0.45\linewidth}
        \includegraphics[width = \linewidth]{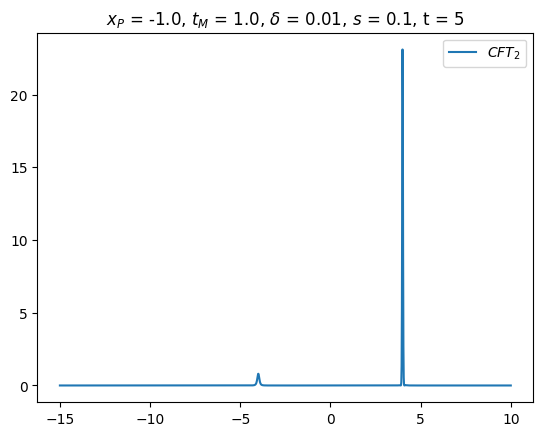}
    \end{minipage}
    \caption{Energy density distribution for boson theory. Here $\beta = \frac{\pi}{2}$ and $p = 1$}
    \label{sec3:boson}
\end{figure}

We have plotted the result for a fixed time $t$ after having inserted both local operator and localized TFD state in fig.\,\ref{sec3:boson} with the background contribution coming from $\langle T \rangle$ and $\langle \bar{T} \rangle$ subtracted.
One can see that the bosonic theory also shares many similar features with holographic theories.
There are two peaks in CFT$_1$ and one big peak in CFT$_2$.
Since the amplitude in CFT$_2$ is much smaller, we also plotted the second CFT separately.
The difference here is the small second peak in CFT\textsubscript{2}.
This can be attributed to the appearance of the moduli derivative term in (\ref{eq.Torus_Ward_Id}).
Indeed, if one subtracts that contribution then the smaller peak in CFT$_2$ will disappear.
This should be interpreted as a local operator dependent background contribution to the energy density.

We have seen that the free theory results are very similar to those for holographic theories, even though the latter are obtained using an approximation.
This justifies the saddle point approximation of correlation functions on a torus by those on a plane, which was used for holographic CFTs.

\section{Evolution of entanglement entropy and mutual information}\label{Sec:EEMI}

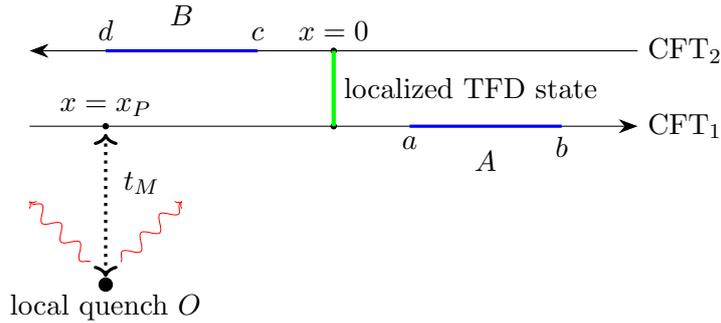
\begin{figure}[tttt]
    \centering
    \begin{tikzpicture}[decoration = snake]
        \draw[{Stealth[length=2.5mm]}-] (-4, 1) -- (4, 1);
        \draw[-{Stealth[length=2.5mm]}] (-4, 0) -- (4, 0);
        \filldraw [black] (0, 0) circle (1pt);
        \filldraw [black] (0, 1) circle (1pt) node[above] {$x = 0$};
        \draw[green, ultra thick] (0, 0) -- (0, 1);
        \node[right] at (0, 0.5) {localized TFD state};
        \node[right] at (4, 0) {CFT$_1$};
        \node[right] at (4, 1) {CFT$_2$};

        \node[above] at (-3,1) {$d$};
        \node[above] at (-1, 1) {$c$};
        \draw[very thick, blue] (-3, 1) -- (-1, 1);
        \node[above] at (-2, 1.2) {$B$};

        \node[below] at (1,0) {$a$};
        \node[below] at (3,0) {$b$};
        \draw[very thick, blue] (1, 0) -- (3, 0);
        \node[below] at (2, -0.2) {$A$};
        
        \filldraw [black] (-3, 0) circle (1pt) node[above] {$x = x_P$};
        \filldraw [black] (-3, -2.1) circle (2.5pt) node[below] {local quench $O$};
        \draw[<->, dotted, very thick] (-3, -2) -- (-3, -0.1);
        \node[right] at (-2.9, -0.75) {$t_M$};
        \draw[->, decorate, red, ultra thin] (-2.8, -1.8) -- (-2, -1);
        \draw[->, decorate, red, ultra thin] (-3.2, -1.8) -- (-4, -1);
    \end{tikzpicture}
    \caption{Our setup of a local operator excitation of the localized TFD state and the computation of entanglement entropy with respect to subsystems $A$ and $B$.}
    \label{mi:setup}
\end{figure}

In this section, we calculate the time evolution of entanglement entropy for different choices of the subsystem and the mutual information between two subsystems within the two-CFT setup.
In particular, we will be interested in calculating the mutual information between two intervals as shown in fig.\,\ref{mi:setup} as the size and position of the intervals are varied.
We will only be concerned with holographic CFTs.
We denote the subsystem chosen in CFT\textsubscript{1} by $A=[a,b]$ and the subsystem chosen in CFT\textsubscript{2} by $B=[c,d]$. This notation will be used through out the whole section \ref{Sec:EEMI}. Moreover, the calculations of entanglement entropy are done by inserting twist operators $\sigma$ and $\bar{\sigma}$ with conformal dimension $h=\bar{h}=\frac{c}{24}(n-\frac{1}{n})$ to the endpoints of intervals $A,B$ and taking $n\to1$ limit of the R\'enyi entropy in the end. We will first analyze the case without the local operator excitation. After that we present the whole results with the local operator. 

\subsection{Contribution by the localized TFD state}\label{Subsec:WH}

We shall first only consider the localized TFD state and analyze its contribution to entanglement entropy and mutual information between CFT\textsubscript{1} and CFT\textsubscript{2}. Since this can be viewed as a mixed-state excitation when one of the CFTs are traced out, we shall denote the entanglement entropy (of subsystem $A$) and mutual information (between subsystems $A$ and $B$) of the localized TFD state by $S^{\text{mix}}_A$ and $I^{\text{mix}}_{AB}$ respectively. 
The additional contribution due to the local operator insertion will be considered in the following subsections.

\subsubsection{Single-interval subsystems}
For the localized TFD state, all the nontrivial two-point functions of the twist operators are 
\begin{align}
    \langle\sigma_1(a)\,\bar{\sigma}_1(b)\rangle
    &=\left\{\frac{4\beta^2}{\pi^2\,\epsilon_a\epsilon_b}
    \sinh\frac{\pi(\Theta_b-\Theta_a)}{2\beta}
    \sinh\frac{\pi(\bar{\Theta}_b-\bar{\Theta}_a)}{2\beta}
    \right\}^{-2h} \\
    \langle\sigma_2(c)\,\bar{\sigma}_2(d)\rangle
    &=\left\{\frac{4\beta^2}{\pi^2\,\epsilon_c\epsilon_d}
    \sinh\frac{\pi(\Theta_d-\Theta_c)}{2\beta}
    \sinh\frac{\pi(\bar{\Theta}_d-\bar{\Theta}_c)}{2\beta}
    \right\}^{-2h} \\
    \langle\sigma_1(a)\,\bar{\sigma}_2(d)\rangle
    &=\left\{\frac{4\beta^2}{\pi^2\,\epsilon_a\epsilon_d}
    \cosh\frac{\pi(\Theta_a+\Theta_d)}{2\beta}
    \cosh\frac{\pi(\bar{\Theta}_a+\bar{\Theta}_d)}{2\beta}
    \right\}^{-2h} \\
    \langle\sigma_2(c)\,\bar{\sigma}_1(b)\rangle
    &=\left\{\frac{4\beta^2}{\pi^2\,\epsilon_c\epsilon_b}
    \cosh\frac{\pi(\Theta_c+\Theta_b)}{2\beta}
    \cosh\frac{\pi(\bar{\Theta}_c+\bar{\Theta}_b)}{2\beta}
    \right\}^{-2h},
\end{align}  
where we set
\be
\epsilon_x=\frac{2s}{\sqrt{((x-t+t_M)^2+s^2)((x+t-t_M)^2+s^2)}}\epsilon,
\ee
and $\Theta_i=\Im w_i,\,\bar{\Theta}_i=-\Im\bar{w}_i\,(i=a,b,c,d)$. We have used indices $1,2$ to denote twist operators in CFT\textsubscript{1,2} respectively, though the twist operators in the two CFTs are identical and can couple to each other. Then the entanglement entropy of subsystems $A$ and $B$ are respectively given by
\begin{align}
    &S^\text{mix}_A=\lim_{n\to1}\frac{1}{1-n}\log\,
    \langle\sigma_1(a)\,\bar{\sigma}_1(b)\rangle \\
    &S^\text{mix}_B=\lim_{n\to1}\frac{1}{1-n}\log\,
    \langle\sigma_2(c)\,\bar{\sigma}_2(d)\rangle\,.
\end{align}

\begin{figure}[ttt]
\centering
     \begin{minipage}{0.40\linewidth}
         \centering
         \includegraphics[width=\linewidth]{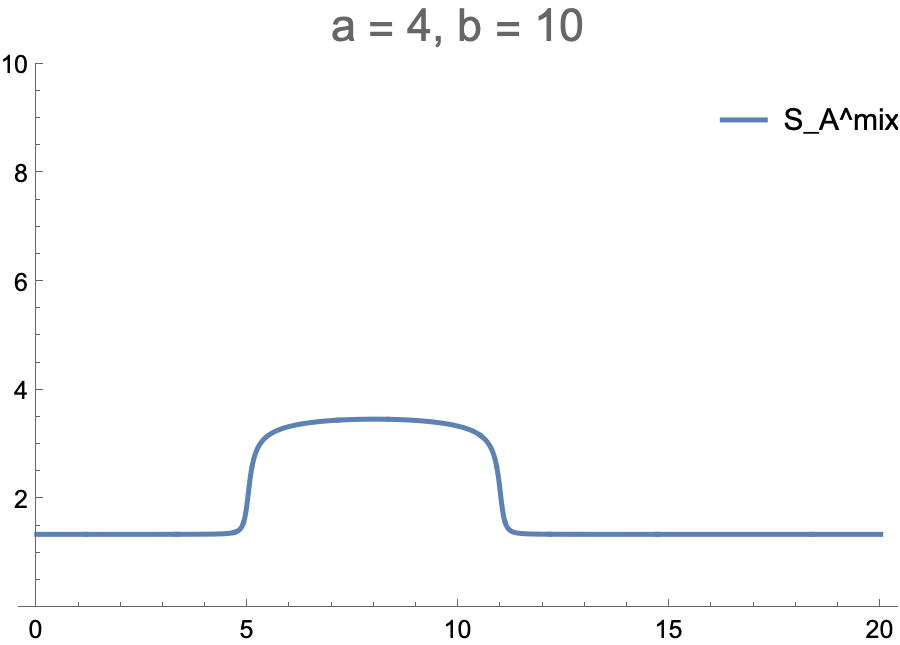}
         \caption{\footnotesize Time evolution of $S_A^{\text{mix}}$ when $A$ is chosen not to include the local operator insertion.}
         \label{S_A^mix(1)}
     \end{minipage}
     \hspace{1cm}
     \begin{minipage}{0.40\linewidth}
         \centering
         \includegraphics[width=\linewidth]{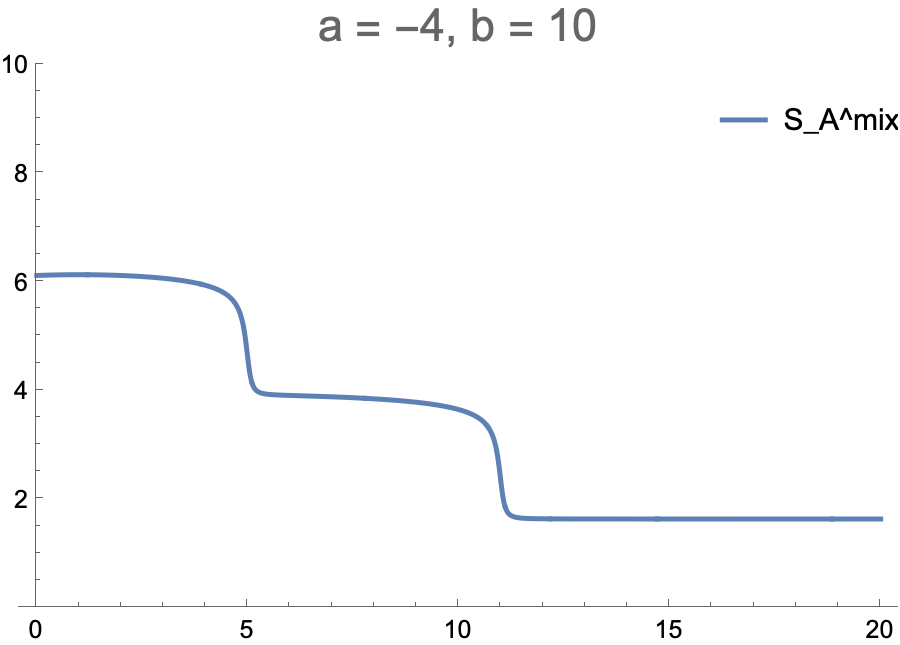}
         \caption{\footnotesize Time evolution of $S_A^{\text{mix}}$ when $A$ is chosen to include the local operator insertion.}
         \label{S_A^mix(2)}
     \end{minipage}
\end{figure}

Without the insertion of a local operator, the physics is symmetric in the two CFTs and the behaviors of entanglement entropy in $A$ and $B$ are identical, so it is sufficient to only analyze $S_A^{\text{mix}}$. As we explain below, we observed that the behaviors of $S_A^{\text{mix}}$ are different in the case where $A$ includes the insertion of localized TFD state from the case where $A$ does not. The time evolution of $S_A^{\text{mix}}$ for these two cases is shown in fig.\ref{S_A^mix(1)} and fig.\ref{S_A^mix(2)} with parameters $t_M=1,\,a=4,\,b=10,\,\delta=0.01,\,s=0.1,\,\epsilon=0.1,\,\beta=1$ and central charge taken to be 1\footnote{While we are working in the approximation where the central charge is sufficiently large, the central charge only appears as an overall scaling factor in the final result. We have therefore set it to be 1 for convenience.}.

\begin{figure}
    \centering
    \includegraphics[width=6cm]{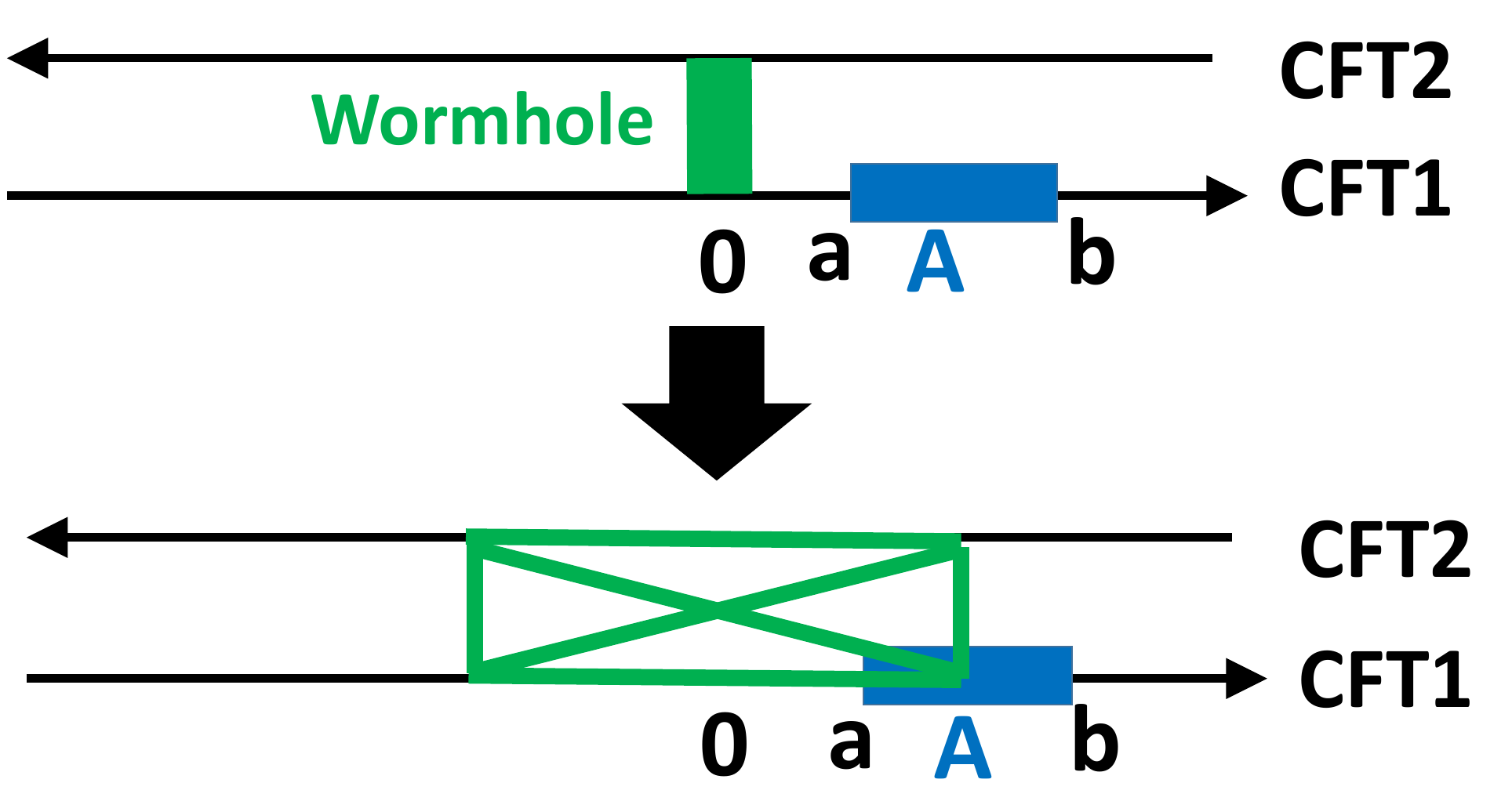}
    \hspace{5mm}
     \includegraphics[width=6cm]{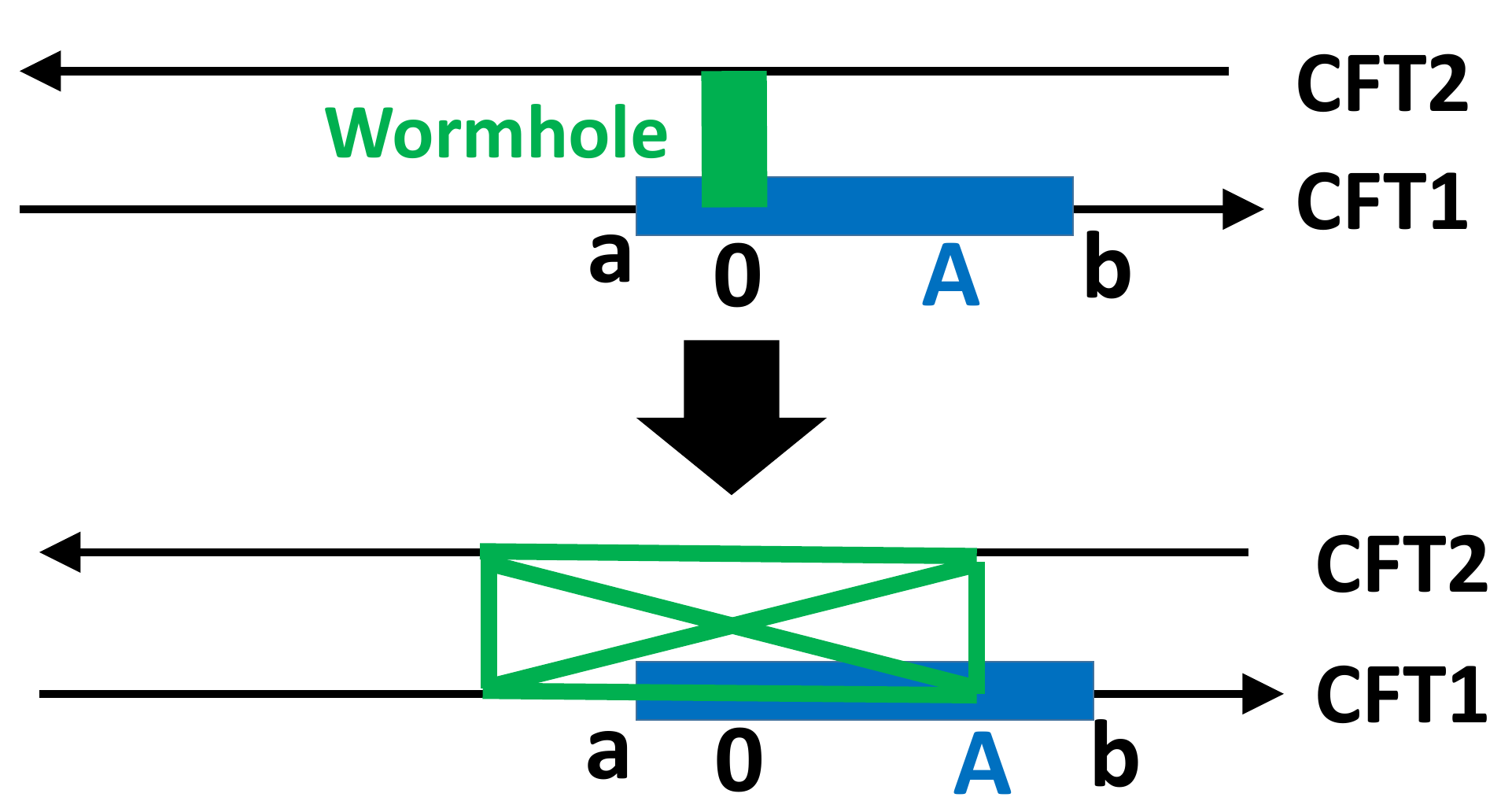}
    \caption{The evolution of quantum entanglement of the localized TFD state, dual to the AdS wormhole, between two CFTs from time $t_M$ to time $t$. The left and right panel show the setups corresponding to fig.\,\ref{S_A^mix(1)} and fig.\,\ref{S_A^mix(2)}, respectively. The green lines describe the bipartite entanglement between four points (\ref{fourp}), which travel at the speed of light.}
    \label{propamix}
\end{figure}

When the insertion is outside subsystem $A$, that is $0\notin[a,b]$ (let us take $a,b>0$ without loss of generality), as plotted in fig.\,\ref{S_A^mix(1)}, a bump appears in $S_A^{\text{mix}}$ in the time interval $a+t_M\leq t\leq b+t_M$. This behavior simply reflects the fact that the localized TFD state initially creates strong entanglement between points $x=0$ in CFT$_1$ and $x=0$ in CFT$_2$ at time $t=t_M$. Under time evolution, each of the entangled points separates into left and right-moving modes propagating at the speed of light, as depicted in the left panel of fig.\,\ref{propamix}. This is the bipartite entanglement between any pair of $P_1,Q_1,P_2$ and $Q_2$, which we have already explained in (\ref{fourp}).
Each entangled pair is expected to be dual to the AdS wormhole with width $\sim O(s)$ because they are regularized by the UV cutoff $s$. 
In this way we can explain the evolution in fig.\,\ref{S_A^mix(1)}.
One can also work out the entanglement entropy analytically in the $s\to 0$ limit ($\epsilon$ is the UV cutoff):
\begin{equation}
    S_A^{\text{mix}}\approx\left\{
    \begin{aligned}
        &\frac{c}{6}\log\left[\frac{\beta}{4\pi s\epsilon^2}
        \left(b-a\right)^3
        \sinh\left(\frac{\pi^2}{\beta}\right) 
        \right] 
        &,&\quad a<t-t_M<b \\[3pt]
        &\frac{c}{3}\log\frac{b-a}{\epsilon} 
        &,&\quad\text{otherwise}
    \end{aligned}\right.
\end{equation}

When the insertion is inside subsystem $A$, that is $0\in[a,b]$ (let us take $|a|<|b|$ without loss of generality), as plotted in fig.\,\ref{S_A^mix(2)}, the behavior of $S_A^{\text{mix}}$ looks like a double-step function. It takes the maximal value at $t=t_M$ and decreases monotonically. There are two distinct drops in $S_A^{\text{mix}}$, located at $t=t_M+|a|$ and $t=t_M+|b|$. This behavior is consistent with what we have discussed above --- that four entangled points $P_1,Q_1,P_2$ and $Q_2$, given by (\ref{fourp}), are propagating at the speed of light, as depicted in the right panel of fig.\,\ref{propamix}. Initially, pairs, $P_1P_2$ and $Q_1Q_2$, are inside the subsystem and we simply observe the contribution to entanglement entropy from both of them. As time evolves, the left-moving mode exits subsystem $A$ first at $t=t_M+|a|$ and we lose its contribution, causing the first drop in fig.\,\ref{S_A^mix(2)}. After that, the right-moving mode also exits $A$, which explains the second drop at $t=t_M+|b|$. The subsequent value of entanglement entropy is the same as that of the vacuum. We can also obtain the analytical result in the small $s$ limit again:
\begin{equation}
    S_A^{\text{mix}}\approx\left\{
    \begin{aligned}
        &\frac{c}{3}\log\left[\frac{\beta\,|ab|}{\pi s\epsilon}
        \sinh\left(\frac{\pi^2}{\beta}\right)
        \right]
        &,&\quad 0<t-t_M<-a \\[3pt]
        &\frac{c}{6}\log\left[\frac{\beta\,|ab|}{\pi s\epsilon^2}
        \left(b-a\right)
        \sinh\left(\frac{\pi^2}{\beta}\right) 
        \right] 
        &,&\quad -a<t-t_M<b \\[3pt]
        &\frac{c}{3}\log\frac{b-a}{\epsilon} 
        &,&\quad t-t_M>b
    \end{aligned}\right.
\end{equation}

\subsubsection{Double-interval subsystems and mutual information}

When calculating the entanglement entropy of subsystem $AB(\equiv A\cup B)$, one encounters the four-point function of twist operators. In holographic CFTs, using the generalized free approximation, we can evaluate this as follows 
\begin{equation}
    \langle\sigma_1\,\bar{\sigma}_1\,\sigma_2\,\bar{\sigma}_2
    \rangle=\langle\sigma_1\,\bar{\sigma}_1\rangle
    \langle\sigma_2\,\bar{\sigma}_2\rangle
    +\langle\sigma_1\,\bar{\sigma}_2\rangle
    \langle\sigma_2\,\bar{\sigma}_1\rangle\,.
\end{equation}
Since our purpose is to calculate the entanglement entropy, it is sufficient to pick up only the dominant term as we take the large central charge limit:
\begin{align}\label{S_AB^mix}
    S^\text{mix}_{AB}&={\min}
    \big\{\lim_{n\to1}\frac{1}{1-n}\log\langle\sigma_1\bar{\sigma}_1\rangle
    \langle\sigma_2\bar{\sigma}_2\rangle\,,\,
    \lim_{n\to1}\frac{1}{1-n}\log\langle\sigma_1\bar{\sigma}_2\rangle
    \langle\sigma_2\bar{\sigma}_1\rangle\big\} \nn\\
    &={\min}\big\{
    S^{\text{mix},\,\text{dis}}_{AB},\,S^{\text{mix},\,\text{con}}_{AB}
    \big\}.
\end{align}
We shall call the first term the ``disconnected phase" and the second term the ``connected phase". These terminologies come from gravity-side calculations. The holographic entanglement entropy is computed by calculating the length of the geodesic whose endpoints coincide with the locations of the twist operators \cite{RT}. In our case, the gravitational dual is given by the BTZ black hole with CFT\textsubscript{1} and CFT\textsubscript{2} lying at the two asymptotic boundaries. The two-point function $\langle\sigma_1\bar{\sigma}_1\rangle$ is now replaced by the geodesic anchored on one of the boundaries, connecting points $x=a$ and $x=b$ lying in CFT\textsubscript{1}. In contrast, $\langle\sigma_1\bar{\sigma}_2\rangle$ should be replaced by the geodesic connecting both of the two boundaries. In this way, we see that the first term in (\ref{S_AB^mix}) corresponds to the case where the geodesics do not connect subsystems $A$ and $B$, while the second term correspond to two geodesics connecting subsystems $A$ and $B$.

The mutual information is defined by
\begin{align}\label{mixed_state_mutual_info}
    I^\text{mix}_{AB}&=S^\text{mix}_A+S^\text{mix}_B-S^\text{mix}_{AB}\nn\\
    &=\max\big\{S^{\text{mix},\,\text{dis}}_{AB}-S^{\text{mix},\,\text{con}}_{AB},0\big\}.
\end{align}
It is clear that in the disconnected phase, the mutual information vanishes.
To have nonzero mutual information, 
we need $S_{AB}^{\text{mix}}$ to be in the connected phase.
However, for most choices of the subsystem, the disconnected phase is preferred and $I^\text{mix}_{AB}=0$ throughout the time evolution. The condition for nonzero mutual information during the time evolution is that the insertion of the localized TFD excitation (i.e. $x=0$) lies inside at least one of the subsystems and the two subsystems have overlapping regions, that is $a<0<b$ or $c<0<d$ together with the condition $-A\cap B\neq\emptyset$\footnote{The minus sign is necessary because the spatial directions of the two CFTs are glued in opposite orientations.}. Only when the subsystem is chosen to satisfy these conditions, we see nonzero mutual information as time evolves. Since our final goal is to examine the signals created by the local operator via the analysis of mutual information, here we only list the plots of $S_{AB}^{\text{mix}}$ and $I_{AB}^{\text{mix}}$ for a few choices of the subsystem that capture the physical characteristics of our model. All the following plots are done with parameters $t_M=1$, $s=0.1,\epsilon=0.1,\beta=1$, and the central charge is taken to be 1.
\begin{figure}[h]
\centering
     \begin{minipage}{0.31\linewidth}
         \centering
         \includegraphics[width=\linewidth]{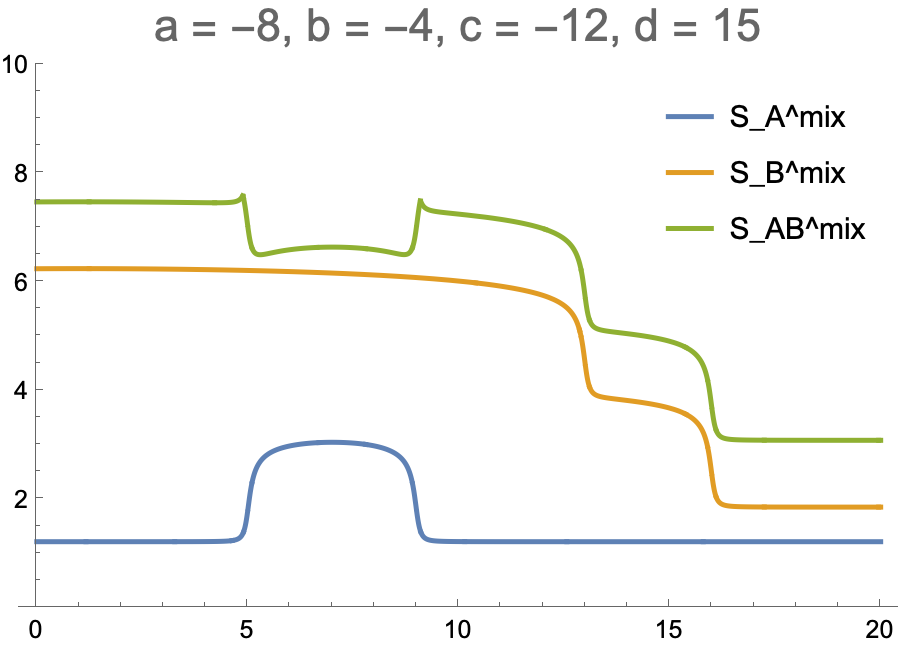}
         \caption{\footnotesize Time evolution of $S^{\text{mix}}_A,$ $S^{\text{mix}}_B$, and $S^{\text{mix}}_{AB}$ for $(a,b,c,d)=(-8,-4,-12,15)$.}
         \label{S_AB^mix(1)}
     \end{minipage}
     \hspace{2mm}
     \begin{minipage}{0.31\linewidth}
         \centering
         \includegraphics[width=\linewidth]{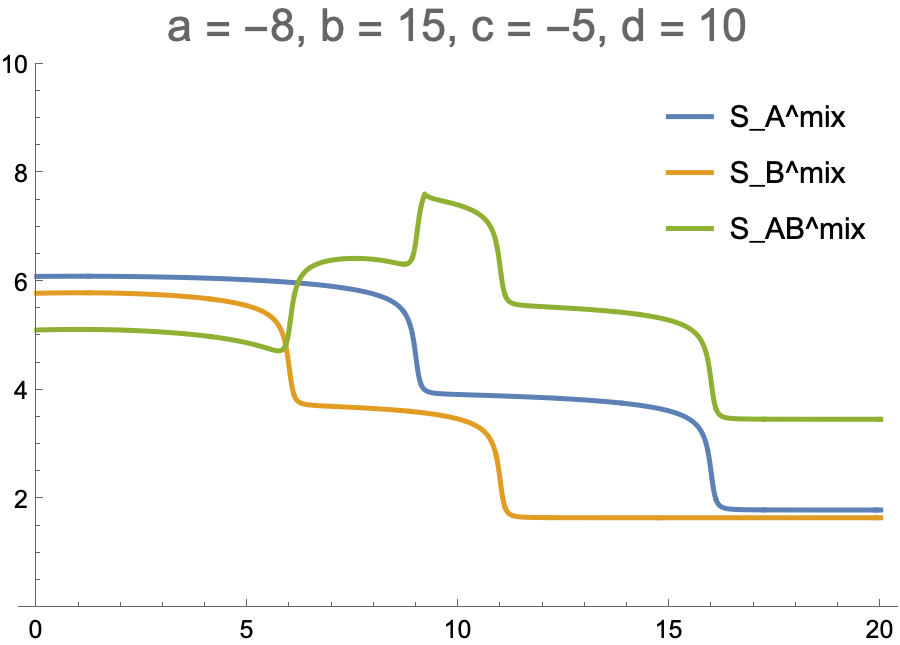}
         \caption{\footnotesize Time evolution of $S^{\text{mix}}_A,$ $S^{\text{mix}}_B$, and $S^{\text{mix}}_{AB}$ for $(a,b,c,d)=(-8,15,-5,10)$.}
         \label{S_AB^mix(2)}
     \end{minipage}
      \hspace{2mm}
     \begin{minipage}{0.31\linewidth}
         \centering
         \includegraphics[width=\linewidth]{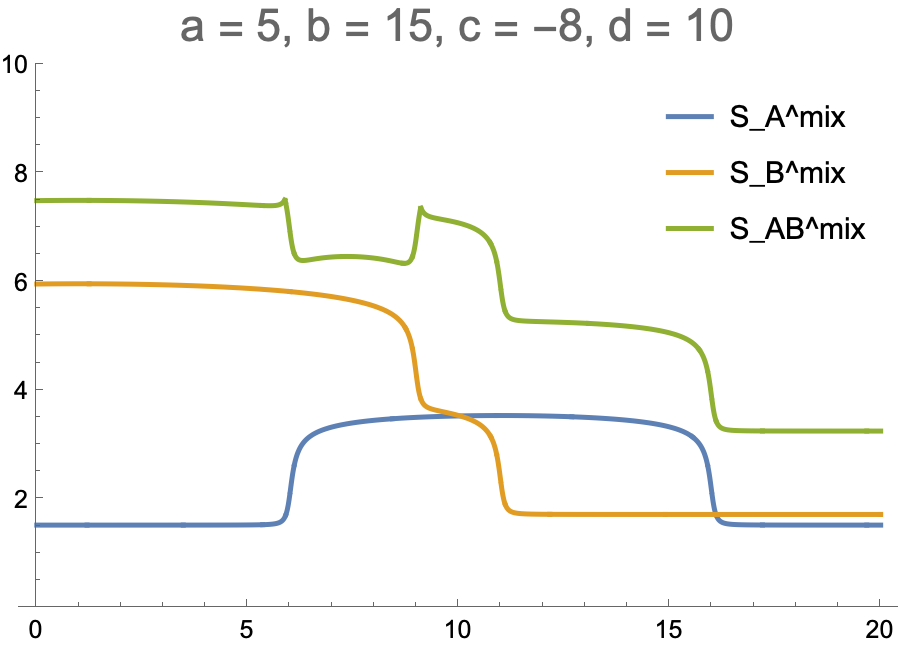}
         \caption{\footnotesize  Time evolution of $S^{\text{mix}}_A,$ $S^{\text{mix}}_B$, and $S^{\text{mix}}_{AB}$ for $(a,b,c,d)=(5,15,-8,10)$.}
         \label{S_AB^mix(3)}
     \end{minipage}
\end{figure}

\begin{figure}[h]
\centering
     \begin{minipage}{0.31\linewidth}
         \centering
         \includegraphics[width=\linewidth]{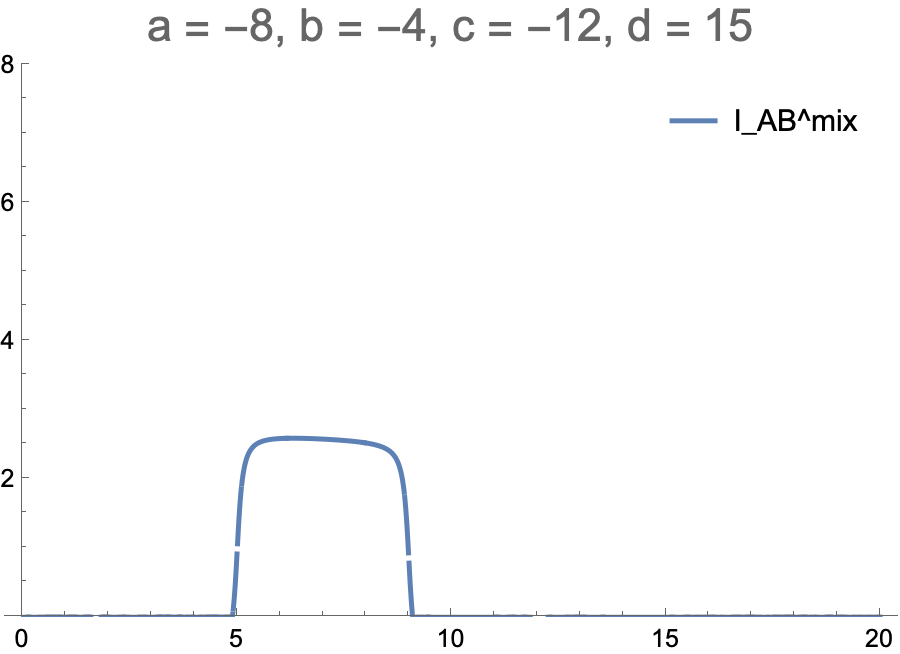}
         \caption{\footnotesize Time evolution of mutual information for the setup of fig.\,\ref{S_AB^mix(1)}.}
         \label{I_AB^mix(1)}
     \end{minipage}
      \hspace{2mm}
     \begin{minipage}{0.31\linewidth}
         \centering
         \includegraphics[width=\linewidth]{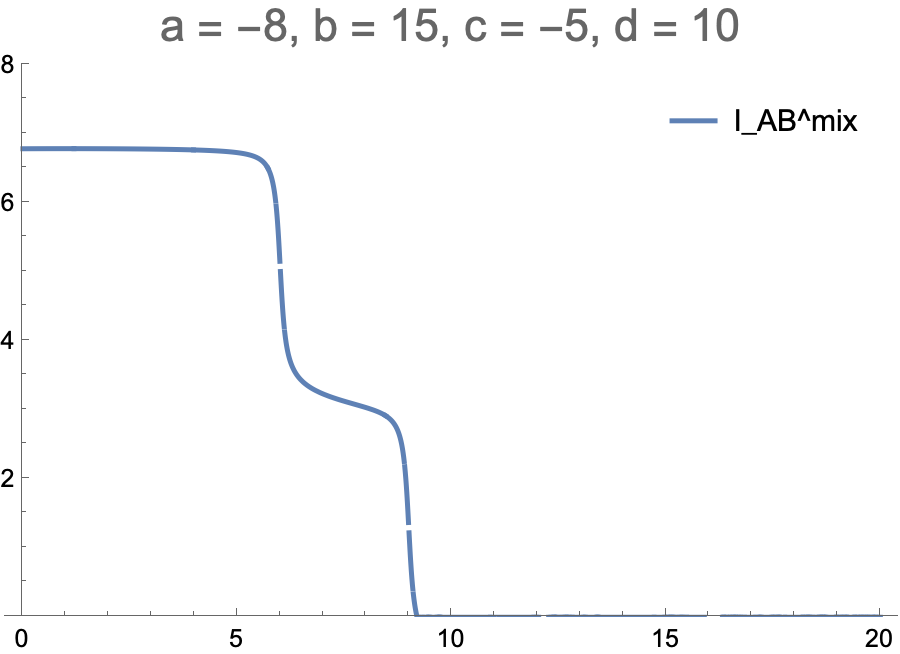}
         \caption{\footnotesize Time evolution of mutual information for the setup of fig.\,\ref{S_AB^mix(2)}.}
         \label{I_AB^mix(2)}
     \end{minipage}
      \hspace{2mm}
     \begin{minipage}{0.31\linewidth}
         \centering
         \includegraphics[width=\linewidth]{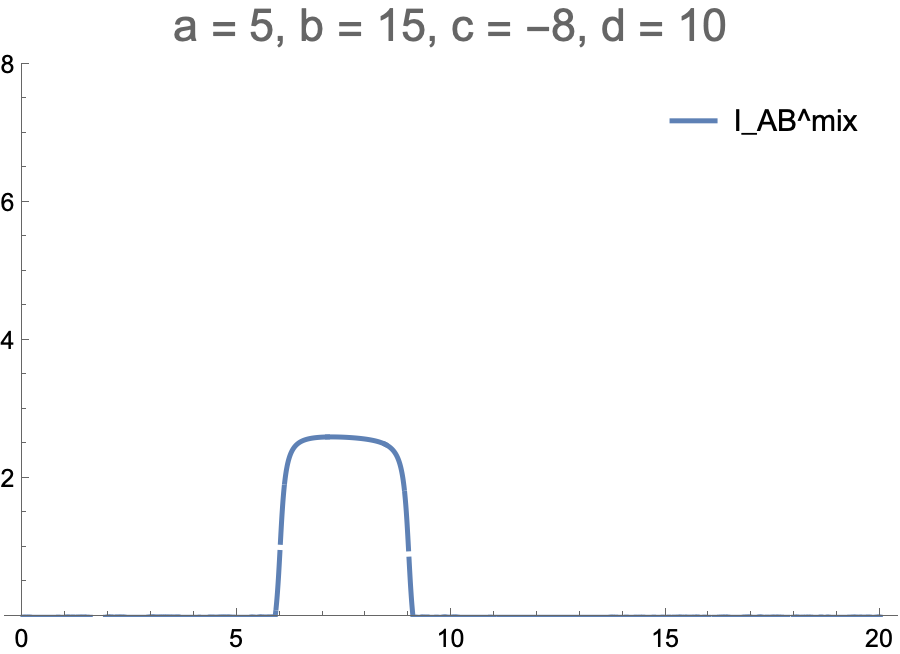}
         \caption{\footnotesize  Time evolution of mutual information for the setup of fig.\,\ref{S_AB^mix(3)}.}
         \label{I_AB^mix(3)}
     \end{minipage}
\end{figure}

\begin{figure}
    \centering
    \includegraphics[width=4.7cm]{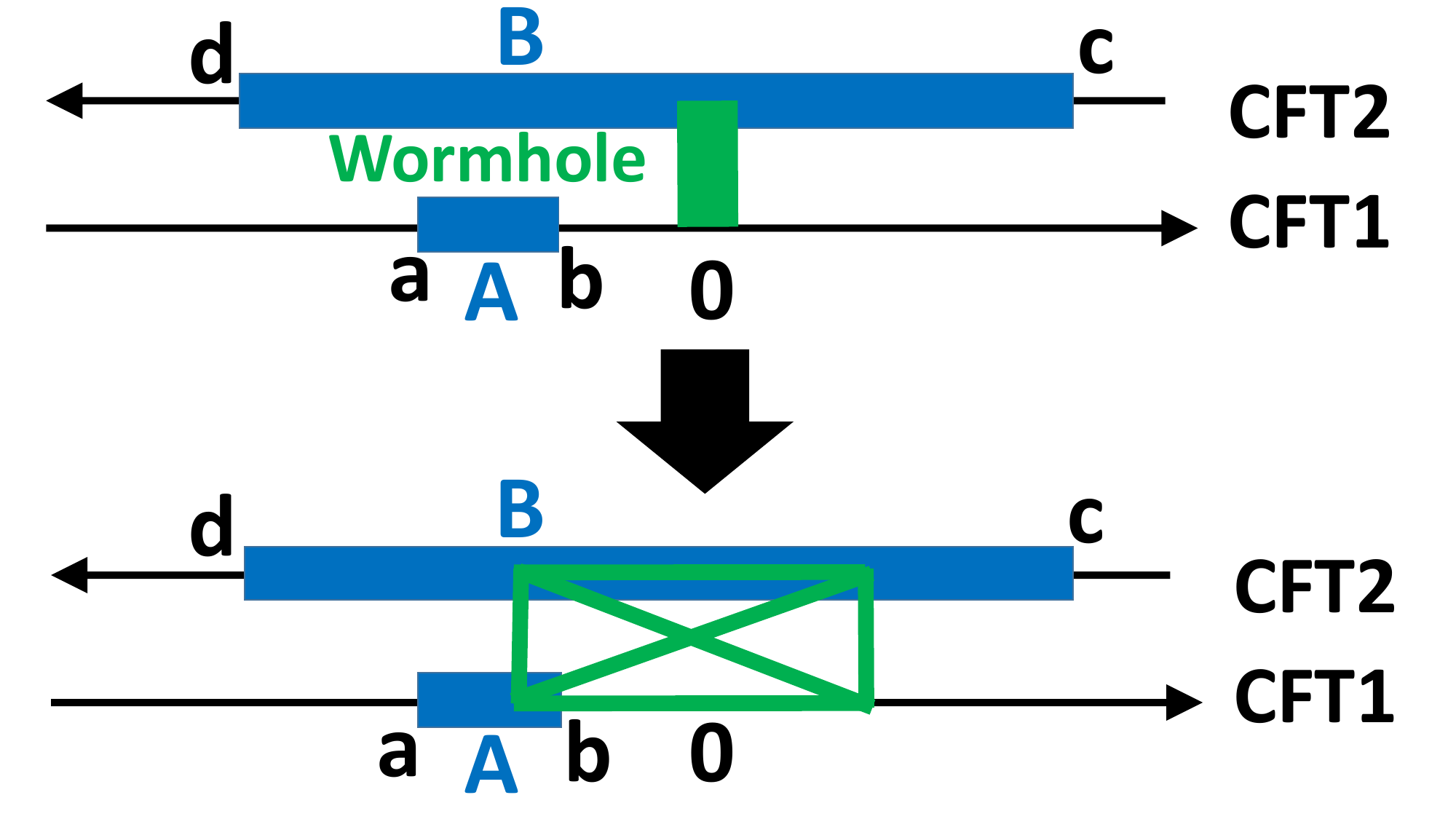}
    \hspace{2mm}
     \includegraphics[width=4.7cm]{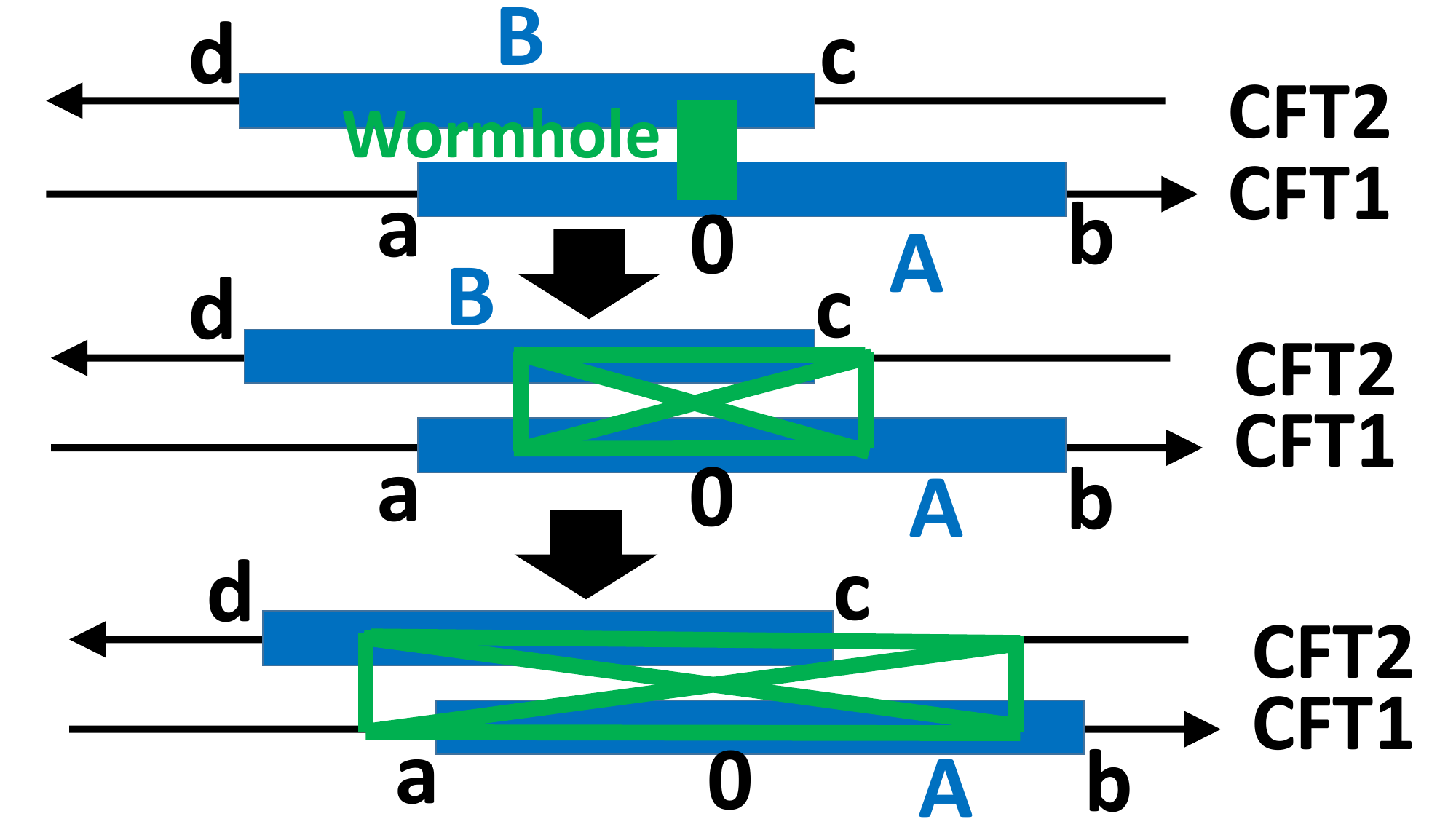}
      \hspace{2mm}
     \includegraphics[width=4.7cm]{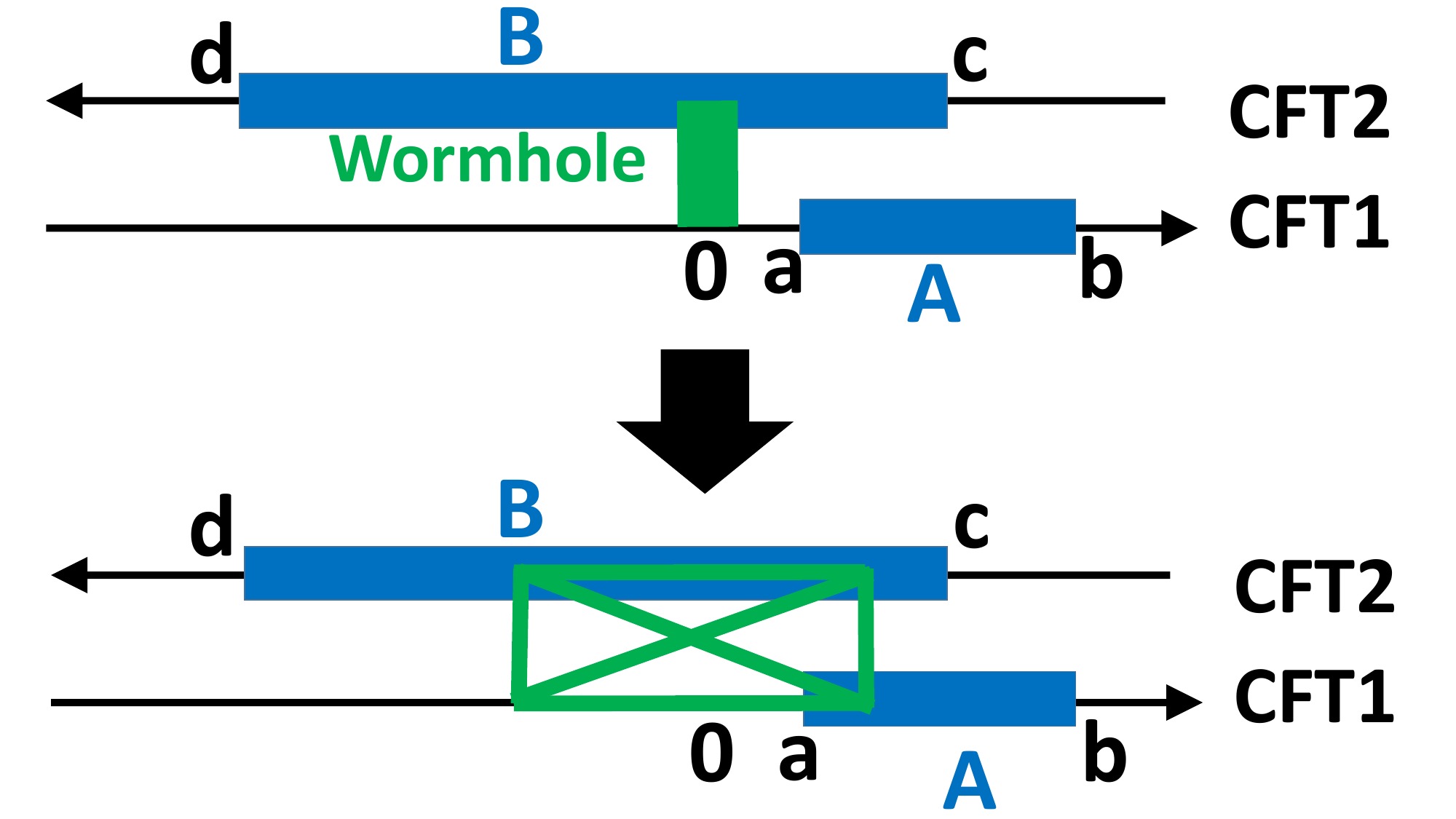}
    \caption{The evolution of quantum entanglement of the localized TFD state, dual to the AdS wormhole, between two CFTs from time $t_M$ to time $t$. The left, middle, and right panel show the setups corresponding to fig.\,\ref{I_AB^mix(1)}, fig.\,\ref{I_AB^mix(2)}, and fig.\,\ref{I_AB^mix(3)}, respectively.
    Green lines show the bipartite entanglement between four points (\ref{fourp}).}
    \label{propaMImix}
\end{figure}

Since the behavior of $S_A^{\text{mix}}$ and $S_B^{\text{mix}}$ was previously discussed, here we only explain the behavior of $S_{AB}^{\text{mix}}$ and $I_{AB}^{\text{mix}}$. For $S_{AB}^{\text{mix}}$, the connected phase, which gives nonzero mutual information, is dominant in time interval $|b|<t-t_M<|a|$ in fig.\,\ref{S_AB^mix(1)}, in time interval $0<t-t_M<|c|$ in fig.\,\ref{S_AB^mix(2)}, and in time interval $|a|<t-t_M<|c|$ in fig.\,\ref{S_AB^mix(3)}. To understand this behavior physically, one has to recall how the four pairwise-entangled modes (\ref{fourp}) propagate, as depicted in the sketches in fig.\,\ref{propaMImix}.

In the cases of fig.\,\ref{S_AB^mix(1)} and fig.\,\ref{S_AB^mix(3)}, the insertion of the localized TFD state is only inside subsystem $B$ and the two subsystems overlap somewhere, but it is only possible for one of the two pairs, $P_1 P_2$ and $Q_1 Q_2$ (\ref{fourp}), to enter that overlapping region as time evolves. Refer to fig.\,\ref{propaMImix} for a sketch. When this does happen, subsystem $AB$ possesses the entangled pair entirely, so the entanglement between subsystem $AB$ and its complement decreases and the mutual information becomes nonzero. After that, the disconnected phase becomes dominant again and the $I_{AB}^{\text{mix}}$ reduces to zero.

On the other hand, in the case of fig.\,\ref{S_AB^mix(2)}, the insertion of the localized TFD state is inside both of the two intervals. Initially, subsystem $AB$ possesses both of the two pairs, $P_1 P_2$ and $Q_1 Q_2$ (\ref{fourp}), and $S_{AB}^{\text{mix}}$ is small while $I_{AB}^{\text{mix}}$ is maximal. It is also equivalent to saying that the connected phase is chosen initially in (\ref{S_AB^mix}). At time $t=t_M+|c|$, one of the two pairs exits subsystem $B$, which causes the first increase in $S_{AB}^{\text{mix}}$ and its corresponding drop in $I_{AB}^{\text{mix}}$. At time $t=t_M+|a|$, though subsystem $AB$ no longer possesses entangled pair and thus the disconnected phase becomes preferred again and the mutual information reduces to zero, half of the endpoints of the entangled modes are inside $AB$ and the others are outside, which enhances $S^{\text{mix}}_{AB}$.

The agreement between the field-theoretic results obtained above and those from gravity-side calculations is obvious. As already discussed earlier in this section, in the gravitational calculation of entanglement entropy, $\langle\sigma_1(a)\bar{\sigma}_1(b)\rangle$ corresponds to the length of a geodesic anchored on points $x=a,b$ in CFT\textsubscript{1} and similarly for $\langle\sigma_2(c)\bar{\sigma}_2(d)\rangle$, while $\langle\sigma_1(a)\bar{\sigma}_2(d)\rangle$ and $\langle\sigma_2(c)\bar{\sigma}_1(b)\rangle$ correspond to the lengths of geodesics connecting the endpoints of subsystems lying on different boundaries. The localized TFD state, which we defined field-theoretically, corresponds holographically to a propagating localized wormhole in Lorentzian signature. As time evolves and the black hole propagates from the boundary into the bulk, it is likely that the geometry around the geodesics changes so that their lengths and corresponding entanglement entropies vary. Especially, for $S_{AB}^{\text{mix}}$, the lengths of connected and disconnected geodesics vary as the wormhole propagates, and the minimal contribution switches back and forth between them. The effects from the localized wormhole will finally dissipate at the asymptotic boundary and the entanglement entropy will coincide with that of the vacuum.

\subsection{Additional contribution by the local operator quench: analytic calculation}\label{Subsec:LO}
We return to our original setup, where we consider a local operator quench $\mathcal{O}$ in the localized TFD state. Our strategy for calculating the entanglement entropy would be to first treat the localized TFD state as a background and to compute the increase in entanglement entropy $\Delta S_A$ (or $\Delta S_B$ for which the same argument holds) relative to the background. Since we have already computed the entanglement entropy of the wormhole background $S^\text{mix}_A$ in the previous subsection, the full contribution to entanglement entropy can simply be expressed as $S_A=S^\text{mix}_A+\Delta S_A$. This strategy is identical to the one taken in \cite{Doi:2025oma}, so readers are encouraged to refer to the paper for further details.

Because the R\'enyi entropy relative to the background can be expressed in terms of correlation functions in the cyclic orbifold theory $\text{CFT}^n/Z_n$ as
\begin{equation}
    \Delta S^{(n)}_A = \frac{1}{1-n}\log\left(\frac{\langle{\mathcal{O}}^{\dagger\otimes n}(X_1,\bar{X}_1)\mathcal{O}^{\otimes n}(X_2,\bar{X}_2)\sigma_n(a) \bar{\sigma}_n(b)\rangle}{\langle{\mathcal{O}}^{\dagger\otimes n}(X_1,\bar{X}_1)\mathcal{O}^{\otimes n}(X_2,\bar{X}_2)\rangle \langle\sigma_n(a)\bar{\sigma}_n(b)\rangle}\right),
\end{equation}
where
\begin{align}
    X_1 &= x_P+i(\delta-it), & \bar{X}_1 &= x_P-i(\delta-it), \\
    X_2 &= x_P-i(\delta+it), & \bar{X}_2 &= x_P+i(\delta+it),
\end{align}
we ought to find out the four-point function $\langle{\mathcal{O}}^{\dagger\otimes n}\mathcal{O}^{\otimes n}\sigma_n \bar{\sigma}_n\rangle$. It turns out we do know how to do this in the Euclidean plane via the heavy-heavy-light-light (HHLL) approximation \cite{Hat,Kap}. This is where the conformal transformation to the annulus and the decompactification of the original toroidal geometry, as described towards the end of section \ref{Subsec:Desc}, plays a role. Working in the $(\zeta,\bar{\zeta})$-coordinates, it can be shown that the functional form of $\Delta S_A$ comes out as
\begin{equation}\label{HHLLformula}
    \Delta S_A = \frac{c}{6}\log\left(\frac{1}{|z|^2}\left|\frac{1-(1-z)^{\alpha_H}}{\alpha_H}\right|^2 |1-z|^{1-\alpha_H}\right),
\end{equation}
where $\alpha_H = \sqrt{1-24h_\mathcal{O}/c}$ and
\begin{equation}
    z=\frac{(\zeta(a)-\zeta(b))(\zeta(w_1)-\zeta(w_2))}{(\zeta(a)-\zeta(w_1))(\zeta(b)-\zeta(w_2))}.
\end{equation}
Furthermore, we can show that $(1-z,1-\bar{z})$ are always of unit length, so --- introducing new variables $(\theta,\bar{\theta})$ defined as
\begin{equation}\label{thetadef}
    1-z = e^{i\theta}, \qquad 1-\bar{z} = e^{-i\bar{\theta}}
\end{equation}
--- (\ref{HHLLformula}) can be further simplified as
\begin{equation}\label{HHLLsimple}
    \Delta S_A = \frac{c}{6}\log\left(\frac{\sin\left(\frac{\alpha_H \theta}{2}\right)\sin\left(\frac{\alpha_H \bar{\theta}}{2}\right)}{\alpha^2_H \sin\left(\frac{\theta}{2}\right)\sin\left(\frac{\bar{\theta}}{2}\right)}\right).
\end{equation}

There is a caveat, however; $\theta,\bar{\theta}$ are actually not well-defined! In other words, there is an ambiguity in the choice of $\theta,\bar{\theta}$. From (\ref{HHLLsimple}), it is clear that the reality of $\Delta S_A$ requires $\theta,\bar{\theta}\in(-2\pi,2\pi)$, and from the definition (\ref{thetadef}), one can, for example, make the replacement $(\theta,\bar{\theta})\mapsto(\theta-2\pi,\bar{\theta}-2\pi)$ if $\theta,\bar{\theta}>0$ and obtain a different $\Delta S_A$. It turns out that there are two physically viable candidates (called `channels' in \cite{Doi:2025oma}) for the pair $(\theta,\bar{\theta})$: one that simultaneously satisfies $\theta,\bar{\theta}>0$ and another that simultaneously satisfies $\theta,\bar{\theta}<0$. The correct prescription for deciding $\Delta S_A$ is to pick the channel that gives the smaller contribution. So (\ref{HHLLsimple}) can be refined to reflect this prescription:
\begin{equation}\label{HHLLfinal}
    \Delta S_A = \min_{(\theta,\bar{\theta})\,\in\,\text{channels}}\left\{\frac{c}{6}\log\left(\frac{\sin\left(\frac{\alpha_H \theta}{2}\right)\sin\left(\frac{\alpha_H \bar{\theta}}{2}\right)}{\alpha^2_H \sin\left(\frac{\theta}{2}\right)\sin\left(\frac{\bar{\theta}}{2}\right)}\right)\right\}.
\end{equation}

In the following analysis, we shall set $x_P = -t_M$ so that the two excitations are always lightlike-separated, where the transmission of entanglement is expected to be the most significant. Single-interval subsystems can either be taken in CFT$_1$ or CFT$_2$, each giving a different contribution to entanglement entropy.

\paragraph{Case I: local excitation $\mathcal{O}$ in CFT$_1$, subsystem $A$ in CFT$_1$.}

This particular setup was studied in \cite{Doi:2025oma}, so we simply cite its results. In terms of the quasiparticle picture, as one of the entangled propagating modes generated by the quench enters and leaves the subsystem, $1-z$ or $1-\bar{z}$ exhibits monodromic behavior. This is captured by nonzero $\theta$ or $\bar{\theta}$ in (\ref{HHLLfinal}), which gives nontrivial $\Delta S_A$.

When the subsystem is taken to the right of the operator insertion, we see a constant bump\footnote{\label{constbump}Technically, a constant bump is only obtained under an approximation where the subsystem is taken to be sufficiently large.} in entanglement entropy relative to the background featuring the localized TFD state. This happens during $a-x_P<t<b-x_P$, when the right-moving mode is inside the subsystem:
\begin{equation}\label{EEpure}
    \Delta S_A = \frac{c}{6} \log \left(\frac{\beta s}{2\pi \alpha_H \delta}\sin(\alpha_H \pi)\tanh\left(\frac{\pi^2}{2\beta}\right)\right).
\end{equation}
When the subsystem is taken to the left of the operator insertion, we once again see a constant bump\textsuperscript{\ref{constbump}} in entanglement entropy, but by a different amount:
\begin{align}\label{EEpurecomp}
    \Delta S_{A} = \frac{c}{6}\log\left(\frac{2t_M}{\alpha_H \delta}\sin(\alpha_H \pi)\right).
\end{align}

\paragraph{Case II: local excitations in CFT$_1$, subsystem $B$ in CFT$_2$.}

When the subsystem is taken to the right of $x=0$, where the localized TFD state was inserted, and in the limit of $0 < c\ll t \ll d$, $z$ comes out to be
\begin{align}    
    z &= \frac{\left(-e^{\frac{\pi^2}{\beta}}+e^{-\frac{\pi^2}{\beta}}\right)\left(\frac{4\pi i\delta}{\beta s}\right)}{\left(-e^{\frac{\pi^2}{\beta}}-1\right)\left(-e^{-\frac{\pi^2}{\beta}}-1\right)} \nonumber \\
    &= -i\frac{4\pi\delta\tanh\left(\frac{\pi^2}{2\beta}\right)}{\beta s} + O(t^{-1}),
\end{align}
and the large $t$ behavior of $\bar{z}$ comes out to be
\begin{align}    
    \bar{z} &= \frac{\left(-e^{-\frac{\pi^2}{\beta}}\left(1+\frac{2\pi s}{\beta t}\right)+e^{-\frac{\pi^2}{\beta}}\right)\left(-e^{-\frac{\pi^2}{\beta}}\frac{\pi i\delta s}{\beta t_M^2}\right)}{\left(-e^{-\frac{\pi^2}{\beta}}-e^{-\frac{\pi^2}{\beta}}\right)\left(-e^{-\frac{\pi^2}{\beta}}-e^{-\frac{\pi^2}{\beta}}\right)} \nonumber \\
    &= i\frac{\pi^2\delta s^2}{2\beta^2 t_M^2 t} + O(t^{-2}).
\end{align}
No monodromic behavior is present for $z$ or $\bar{z}$, but one can still compute the leading order contribution to $\Delta S_B$:
\begin{equation}
    \Delta S_B = \frac{c}{9}(1-\alpha_H^2)\left(\frac{\pi\delta\tanh\left(\frac{\pi^2}{2\beta}\right)}{\beta s}\right)^2,
\end{equation}
which is significantly smaller than what we obtained for case I, where the subsystem was taken in CFT$_1$. This is expected as we have already seen in section \ref{Subsec:EnergyHol} that energy transmission to CFT$_2$ is rather limited.

When the subsystem is taken to the left of $x=0$, and in the limit of $c\ll -t \ll d < 0$, $z$ comes out to be
\begin{align}    
    z &= \frac{\left(-e^{\frac{\pi^2}{\beta}}+e^{\frac{\pi^2}{\beta}}\left(1-\frac{2\pi s}{\beta t}\right)\right)\left(\frac{4\pi i\delta}{\beta s}\right)}{\left(-e^{\frac{\pi^2}{\beta}}-1\right)\left(-e^{\frac{\pi^2}{\beta}}-1\right)} \nonumber \\
    &= -i\frac{2\pi^2\delta}{\beta^2 \cosh^2\left(\frac{\pi^2}{2\beta}\right)t} + O(t^{-2}),
\end{align}
and the large $t$ behavior of $\bar{z}$ comes out to be
\begin{align}    
    \bar{z} &= \frac{\left(-e^{\frac{\pi^2}{\beta}}+e^{-\frac{\pi^2}{\beta}}\right)\left(-e^{-\frac{\pi^2}{\beta}}\frac{\pi i\delta s}{\beta t_M^2}\right)}{\left(-e^{\frac{\pi^2}{\beta}}-e^{-\frac{\pi^2}{\beta}}\right)\left(-e^{-\frac{\pi^2}{\beta}}-e^{-\frac{\pi^2}{\beta}}\right)} \nonumber \\
    &= i\frac{\pi\delta s \tanh\left(\frac{\pi^2}{\beta}\right)}{2\beta t_M^2} + O(t^{-1}).
\end{align}
Again, the leading order contribution to $\Delta S_B$ is given by
\begin{equation}
    \Delta S_B = \frac{c}{576}(1-\alpha_H^2)\left(\frac{\pi\delta s \tanh\left(\frac{\pi^2}{\beta}\right)}{\beta t_M^2}\right)^2.
\end{equation}

When the subsystem is a double interval, as in section \ref{Subsec:WH}, the entanglement entropy is obtained as a sum of two single-interval entanglement entropies. There are two distinct ways of picking the pair of single intervals: the connected and disconnected phases. One must then pick the sum that gives the smaller contribution. Therefore, the entanglement entropy for subsystem $AB$ is given by 
\begin{equation}\label{S_AB}
    S_{AB}={\min}\big\{
    S^{\text{mix},\,\text{dis}}_{AB}+\Delta S_{AB}^{\text{dis}},S^{\text{mix},\,\text{con}}_{AB}+\Delta S_{AB}^{{\text{con}}}
    \big\}
\end{equation}
This can also be derived field-theoretically; refer to appendix \ref{Append:Derivation} for the derivation. The mutual information is then once again computed as $I_{AB}=S_A+S_B-S_{AB}.$

\subsection{Numerical evaluation of entanglement entropy and interpretation}

\begin{figure}[h]
\centering
     \begin{minipage}{0.31\linewidth}
         \centering
         \includegraphics[width=\linewidth]{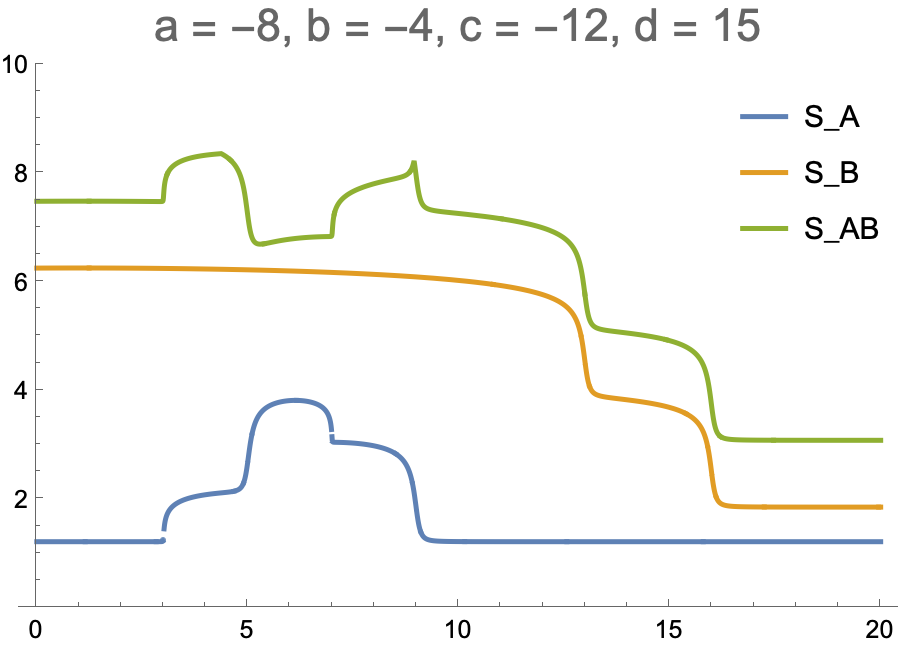}
         \caption{\footnotesize Time evolution of $S_A, S_B$, and $S_{AB}$ in the presence of a local operator for $(a,b,c,d)=(-8,-4,-12,15)$.}
         \label{fig:sab1}
     \end{minipage}
      \hspace{2mm}
     \begin{minipage}{0.31\linewidth}
         \centering
         \includegraphics[width=\linewidth]{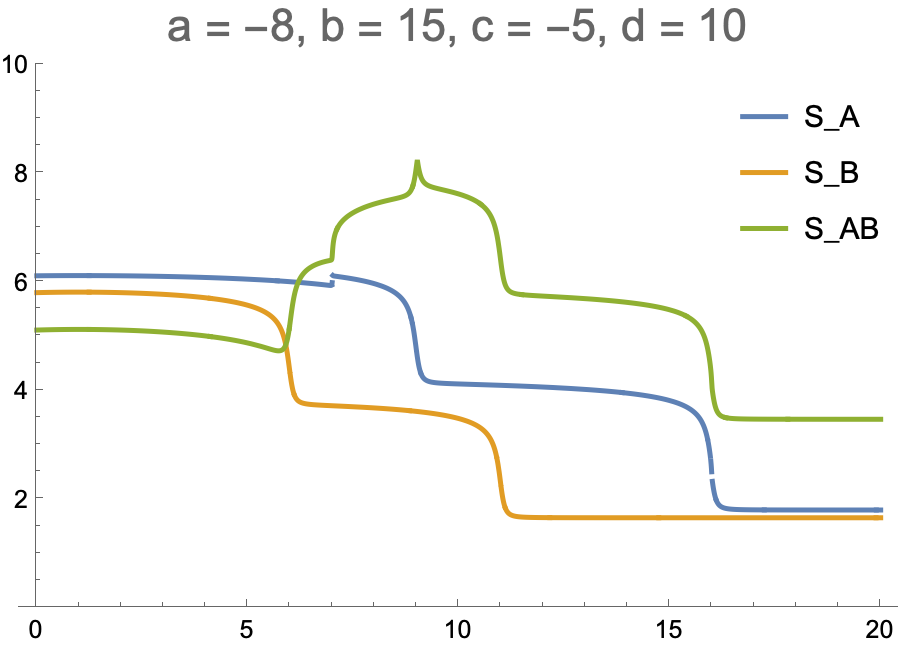}
         \caption{\footnotesize Time evolution of $S_A, S_B$, and $S_{AB}$ in the presence of a local operator for $(a,b,c,d)=(-8,15,-5,10)$.}
         \label{fig:sab2}
     \end{minipage}
      \hspace{2mm}
     \begin{minipage}{0.31\linewidth}
         \centering
         \includegraphics[width=\linewidth]{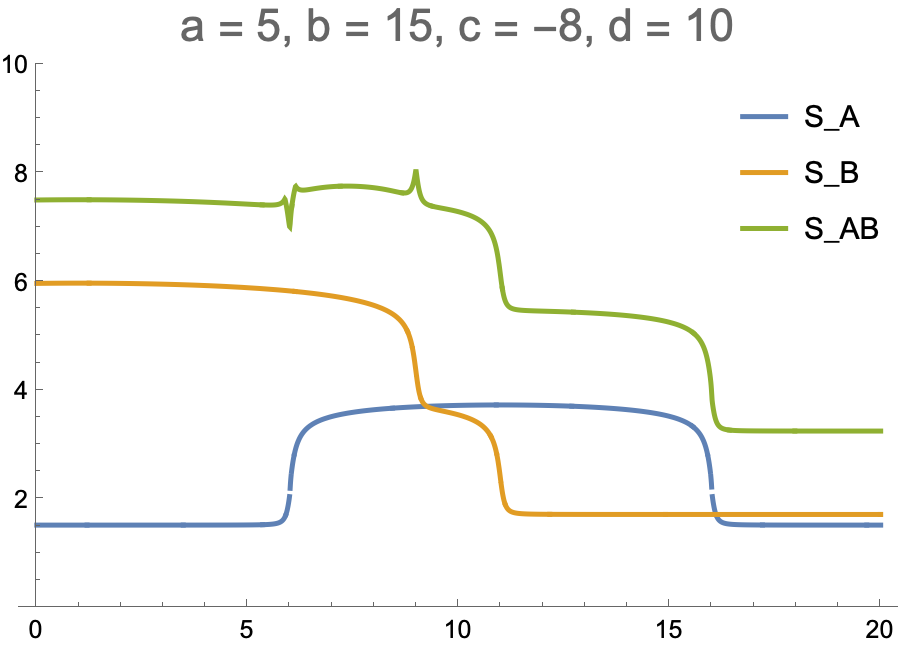}
         \caption{\footnotesize Time evolution of $S_A, S_B$, and $S_{AB}$ in the presence of a local operator for $(a,b,c,d)=(5,15,-8,10)$.}
         \label{fig:sab3}
     \end{minipage}
\end{figure}

\begin{figure}[h]
\centering
     \begin{minipage}{0.31\linewidth}
         \centering
         \includegraphics[width=\linewidth]{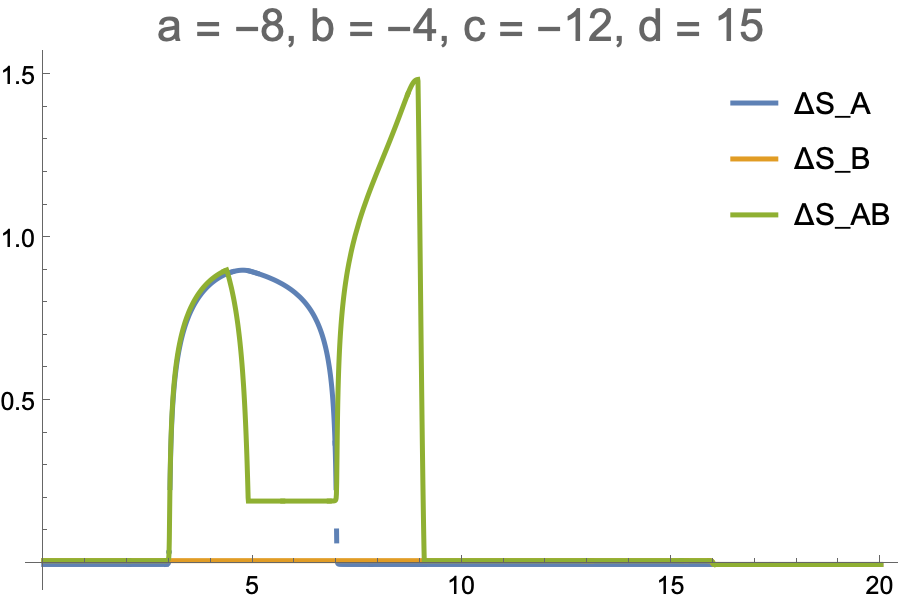}
         \caption{\footnotesize Time evolution of $\Delta S_A, \Delta S_B$, and $\Delta S_{AB}$
         for the setup of fig.\,\ref{fig:sab1}.}
         \label{DeltaS(1)}
     \end{minipage}
      \hspace{2mm}
     \begin{minipage}{0.31\linewidth}
         \centering
         \includegraphics[width=\linewidth]{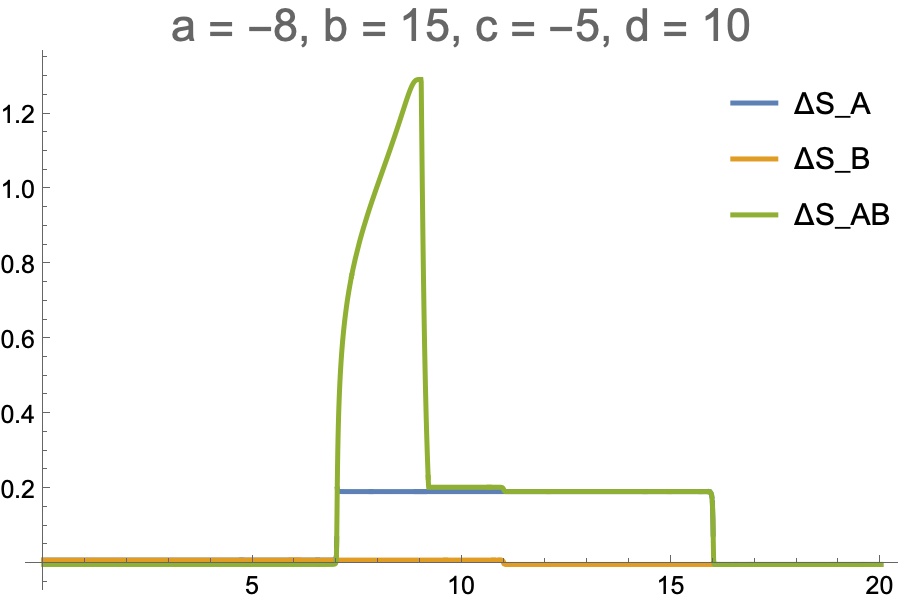}
         \caption{\footnotesize  Time evolution of $\Delta S_A, \Delta S_B$, and $\Delta S_{AB}$
         for the setup of fig.\,\ref{fig:sab2}.}
         \label{DeltaS(2)}
     \end{minipage}
      \hspace{2mm}
     \begin{minipage}{0.31\linewidth}
         \centering
         \includegraphics[width=\linewidth]{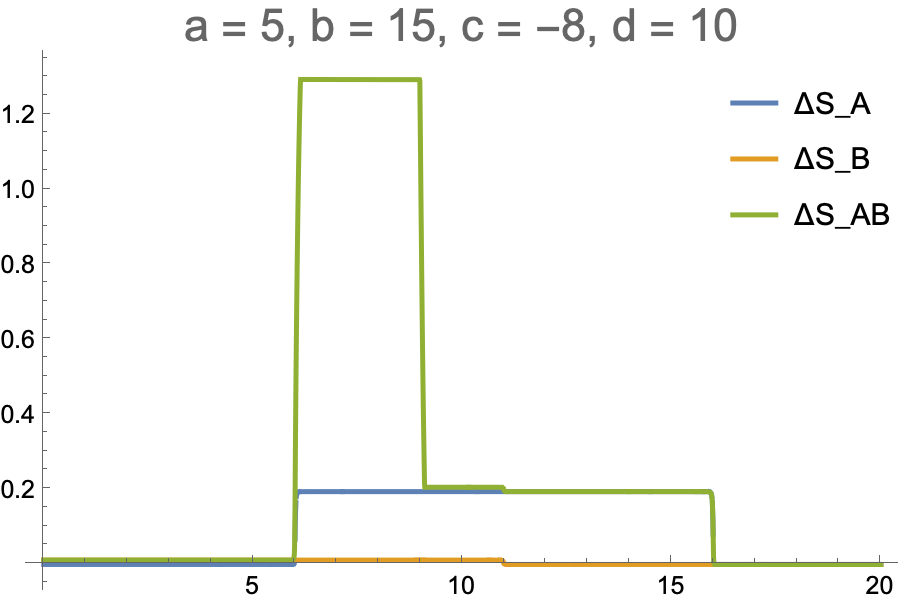}
         \caption{\footnotesize Time evolution of $\Delta S_A, \Delta S_B$, and $\Delta S_{AB}$
         for the setup of fig.\,\ref{fig:sab3}.}
         \label{DeltaS(3)}
     \end{minipage}
\end{figure}

\begin{figure}[h]
\centering
     \begin{minipage}{0.31\linewidth}
         \centering
         \includegraphics[width=\linewidth]{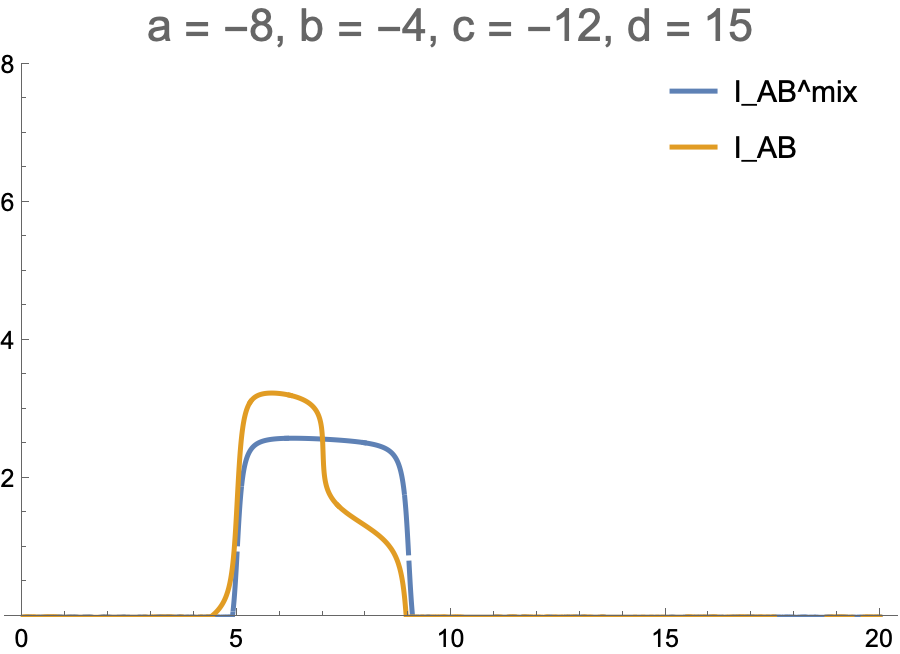}
         \caption{\footnotesize Time evolution of mutual information with and without local operator for the setup of  fig.\,\ref{fig:sab1}. We find enhancement $\Delta I_{AB}>0$ during time interval $|b|<t-t_M<|a|-|x_P|$.}
         \label{I_AB(1)}
     \end{minipage}
      \hspace{2mm}
     \begin{minipage}{0.31\linewidth}
         \centering
         \includegraphics[width=\linewidth]{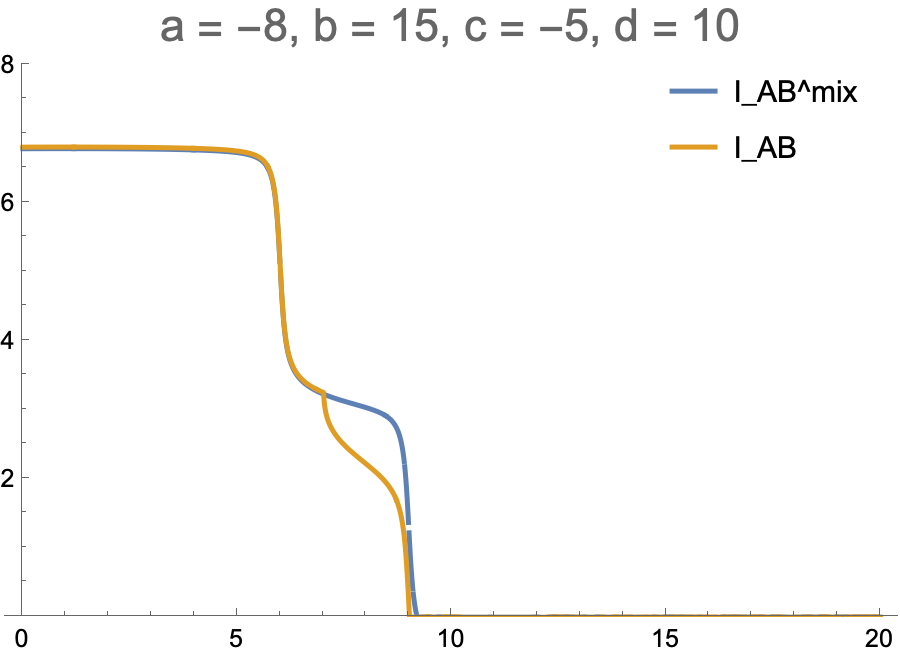}
         \caption{\footnotesize Time evolution of mutual information with and without local operator for the setup of  fig.\,\ref{fig:sab2}. We find that $\Delta I_{AB}$ takes a very small but positive value for $t<|a|+x_P$. For later time we find scrambling behavior $\Delta I_{AB}\leq 0$.}  
         \label{I_AB(2)}
     \end{minipage}
      \hspace{2mm}
     \begin{minipage}{0.31\linewidth}
         \centering
         \includegraphics[width=\linewidth]{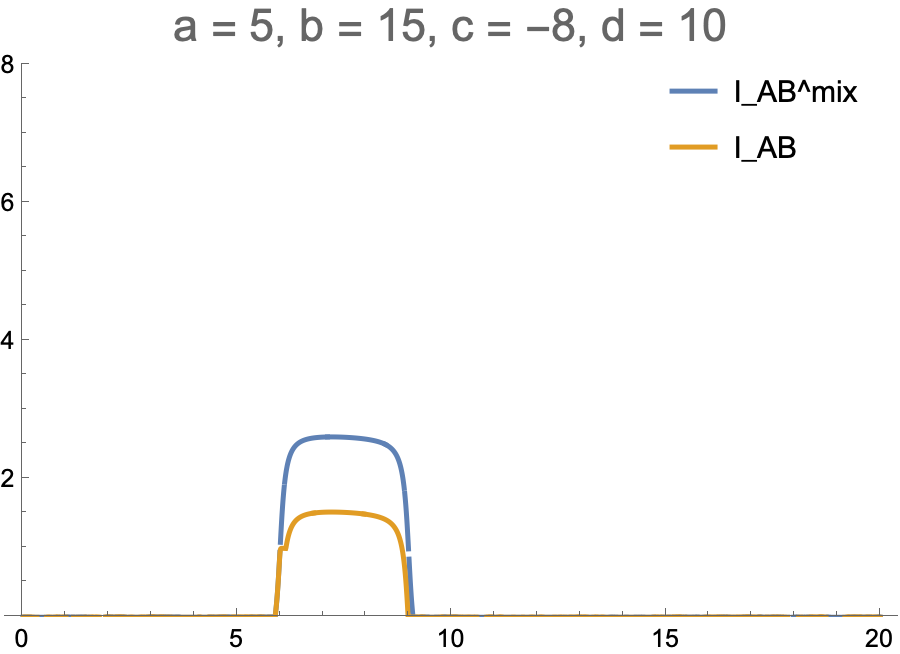}
         \caption{\footnotesize Time evolution of mutual information with and without local operator for the setup of  fig.\,\ref{fig:sab3}. We always find scrambling behavior $\Delta I_{AB}\leq 0$.}  
         \label{I_AB(3)}
     \end{minipage}
\end{figure}

We now compute the entanglement entropy and mutual information for double-interval subsystems $A \cup B$ when the localized TFD state is excited via a local operator insertion. We shall do this numerically by evaluating (\ref{S_AB}).
As seen in section \ref{Subsec:WH} for $S_A$ and $S_B$, one expects a connected-disconnected phase transition from the formula (\ref{S_AB}) and nonzero mutual information when the connected phase is dominant.
We consider the same three subsystems that were studied in section \ref{Subsec:WH} (refer to figs.\,\ref{S_AB^mix(1)}-\ref{I_AB^mix(3)}). The time evolution of $S_A,S_B$ and $S_{AB}$ is plotted in figs.\,\ref{fig:sab1}-\ref{fig:sab3}. The change of entanglement entropy due to the local excitation, i.e. $\Delta S_{X}\equiv S_{X}-S_{X}^{\text{mix}}$ for $X=A,B,AB$, is shown in figs.\,\ref{DeltaS(1)}-\ref{DeltaS(3)}.
Finally the time evolution of mutual information before and after the insertion of local operator is plotted in figs.\,\ref{I_AB(1)}-\ref{I_AB(3)}. We choose $x_P=-1,t_M=1,\delta=0.01, s=0.1, \ep=0.1, \ap_H=0.4, \beta=1$ and take the central charge to be $1$ for all plots.
 
Let us first give a brief summary of our findings before diving into details. It is known that when we add a perturbation to subsystem $A$ of an entangled quantum system $AB$, the correlation between $A$ and $B$ typically gets destroyed and its mutual information decreases under a chaotic time evolution \cite{Hayden:2007cs,Sekino:2008he,Shenker:2013pqa,Maldacena:2015waa,Hosur:2015ylk} --- this is called the scrambling effect. Naively, we can expect the same to happen in our setup when we insert a local operator in the localized TFD state. In fact, this does happen in fig.\,\ref{I_AB(3)}, where quite clearly $\Delta I_{AB}\coloneqq I_{AB}-I_{AB}^{\text{mix}}<0$. However, interestingly, we also observe the opposite effect (may be called the descrambling effect) where the mutual information gets enhanced, i.e. $\Delta I_{AB}>0$. Indeed, in fig.\,\ref{I_AB(1)}, the mutual information $I_{AB}$ in the presence of the local operator excitation is clearly larger than the mutual information $I^{\text{mix}}_{AB}$ without the local excitation for a certain time period. In fig.\,\ref{I_AB(2)}, we also find that $\Delta I_{AB}$ takes a very small but positive value for $t<|a|+x_P$. So how does descrambling happen for certain choices of the subsystem?


\begin{figure}[ttt]
    \centering
    \includegraphics[width=0.45\linewidth]{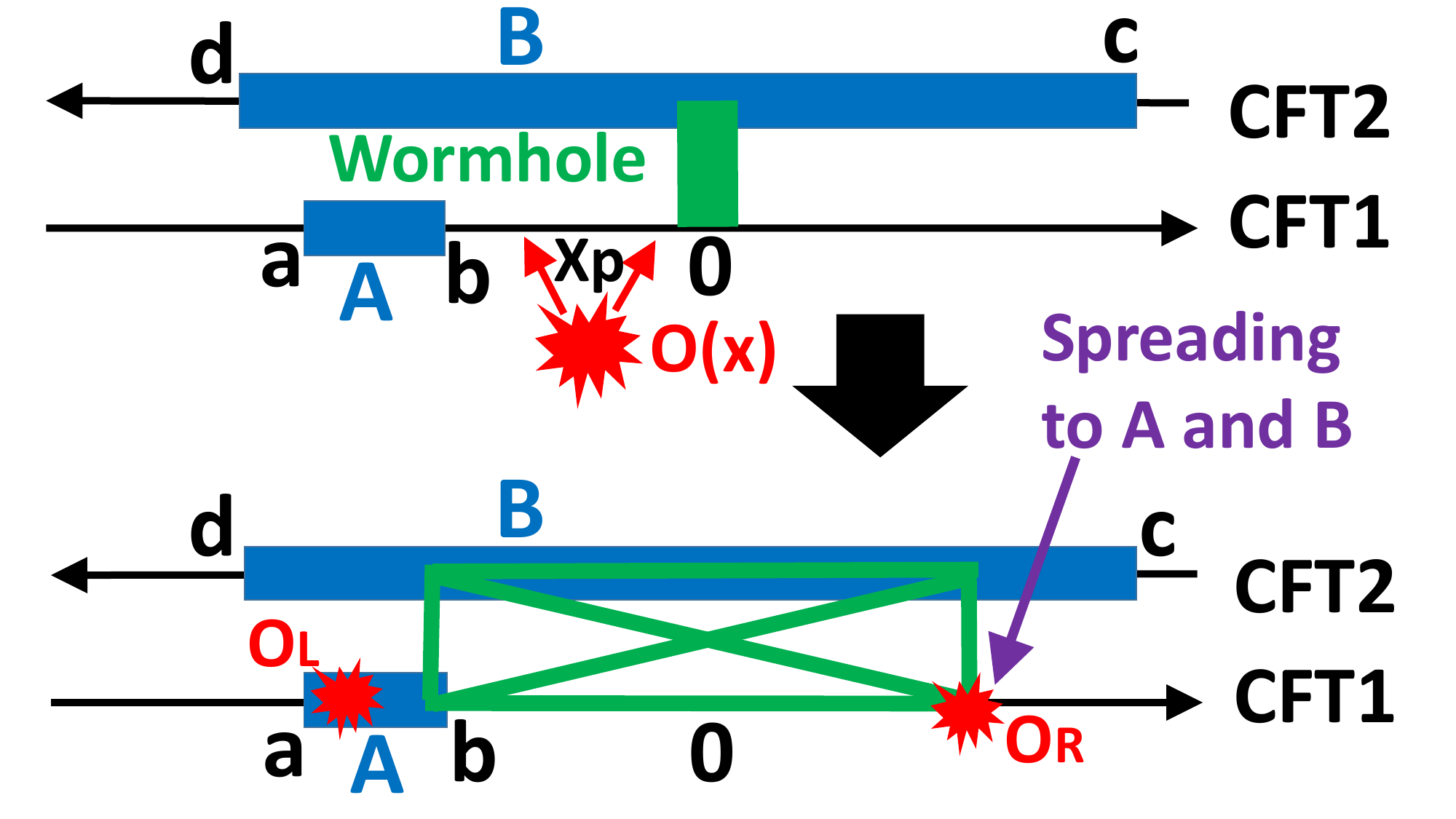}
        \caption{A sketch of the evolution of entanglement for the setup of fig.\,\ref{I_AB(1)}, 
       i.e. $(a,b,c,c)=(-8,-4,-12,15)$. In this case the right-moving mode of the local operator spreads to $B$ (and also a little to $A$) due to the entanglement induced by the wormhole. The left-moving mode is away from the wormhole and is not affected. Thus $S_{AB}$ is not increased by the local operator due to the spreading effect. However, $S_A$ is simply increased by the local operator entanglement. Thus the mutual information is enhanced by the local operator excitation $\Delta I_{AB}> 0$.
     }
    \label{fig:NSCR}
\end{figure}

\begin{figure}[ttt]
    \centering
    \includegraphics[width=0.45\linewidth]{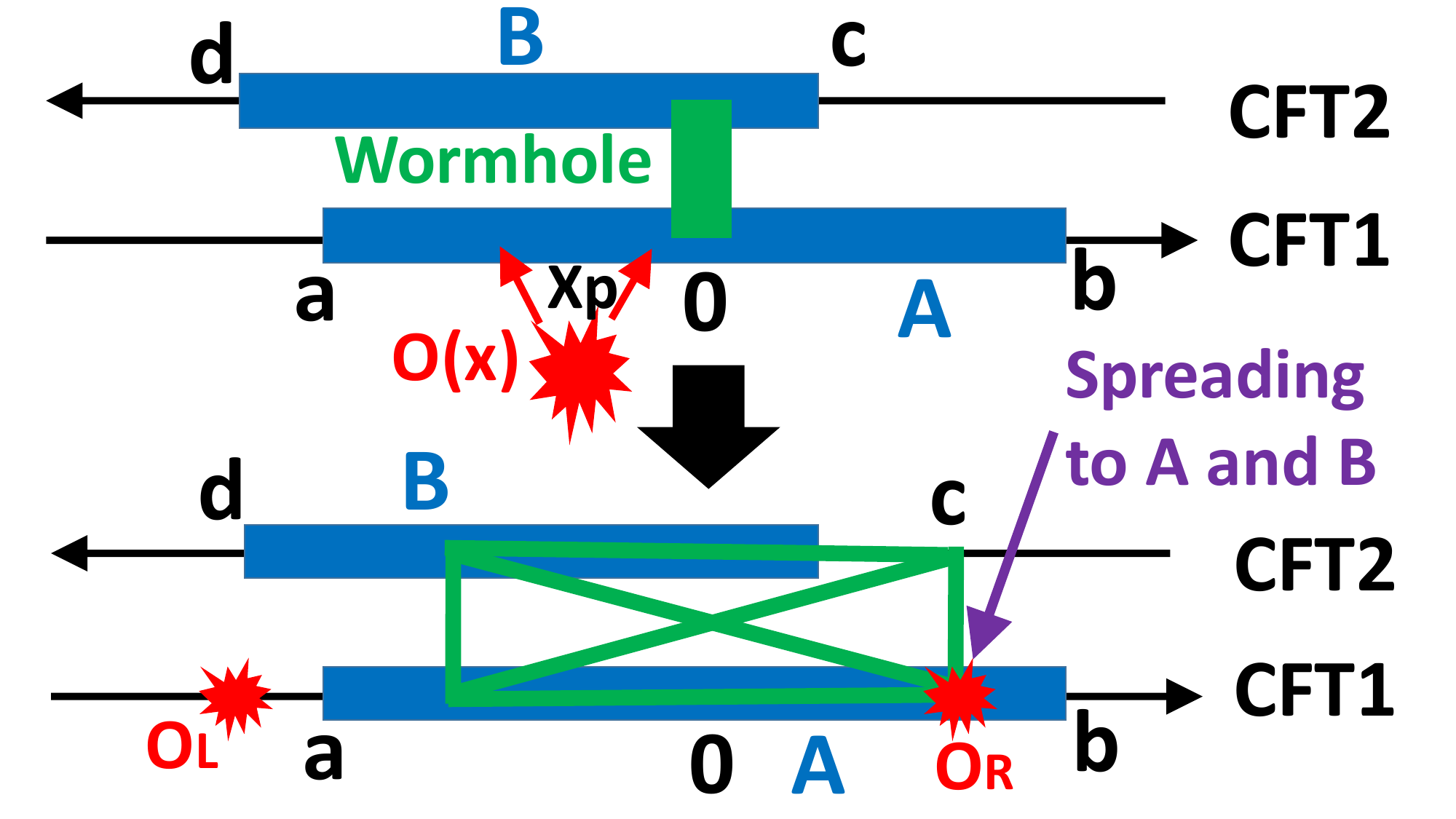}
    \hspace{3mm}
     \includegraphics[width=0.45\linewidth]{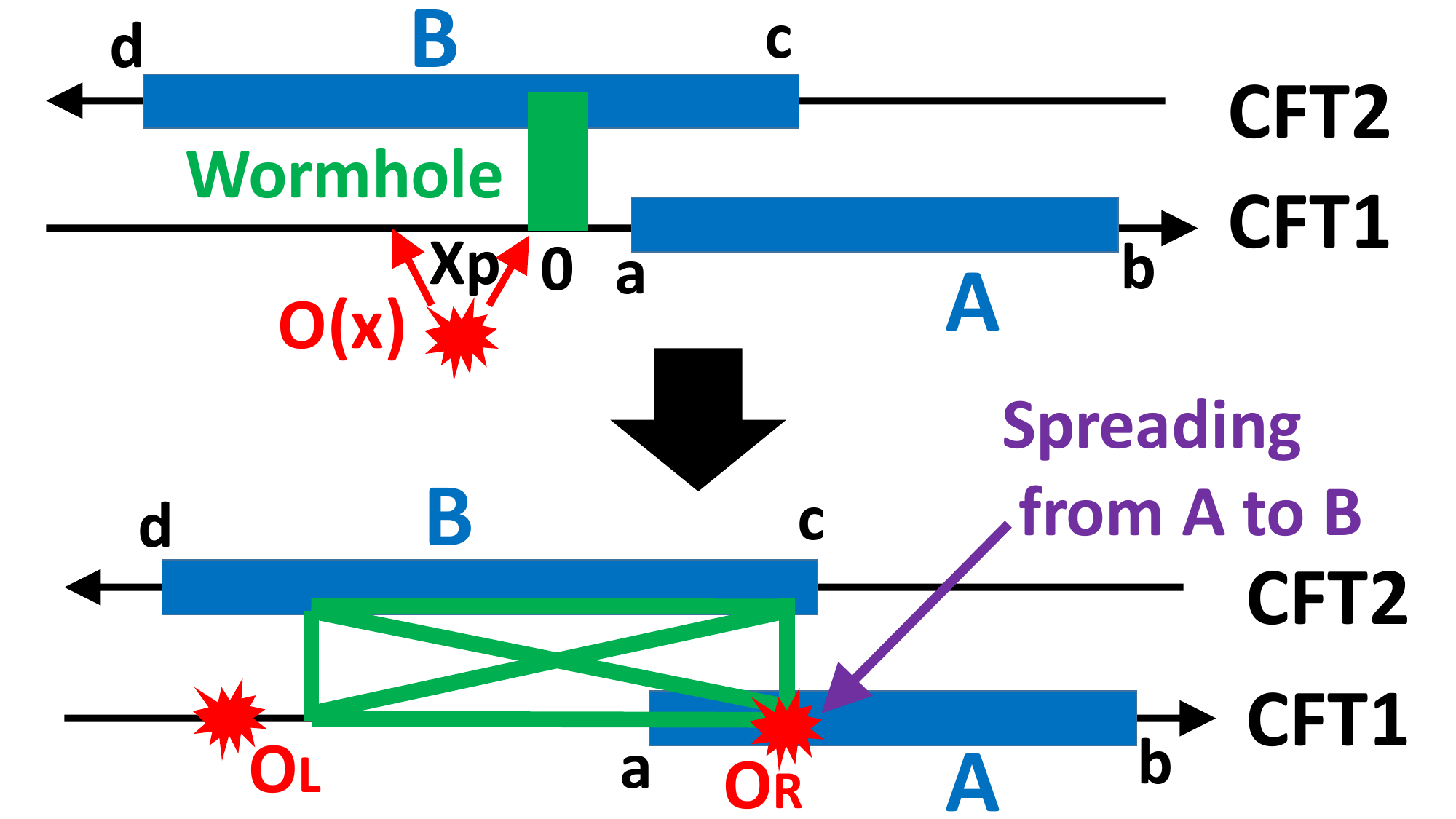}
        \caption{Sketches of the evolution of entanglement for the setups of fig.\,\ref{I_AB(2)} (left panel) and fig.\,\ref{I_AB(3)} (right panel).  They correspond to $(a,b,c,c)=(-8,15,-5,10)$ 
        and $(5,15,-8,10)$, respectively. After the time evolution depicted in the lower sketches, the right-moving mode of the local operator again spreads from $A$ to $B$ due to the entanglement induced by the wormhole. The left-moving mode is away from the wormhole and is not affected. Thus $S_{AB}$ is simply increased by the local operator. However, $S_A$ does not increase by the local operator due to the spreading effect. Thus the mutual information is suppressed $\Delta I_{AB}<0$. Moreover, we would like to note that at early times $t<|a|+x_P$, the setup in the above left panel (i.e. fig.\,\ref{I_AB(2)}), $\Delta I_{AB}$ takes a very small but positive value because the right-moving part of local operator transmits to subsystem $B$ and this adds the correlation between $A$ and $B$.
        }
    \label{fig:SCR}
\end{figure}

It should first be noted that, regardless of the choice of subsystem, $\Delta S_B$ is always small relative to other quantities such as $\Delta S_A$ and $\Delta S_{AB}$. This is expected because we picked the regularization parameters $\delta,s$ to satisfy $\delta/s\ll1$, so the transmission rate from CFT$_1$ to CFT$_2$ is small. Thus the dominant contribution to the change in mutual information $\Delta I_{AB}$ comes from $\Delta S_A$ and $\Delta S_{AB}$ in our setups.

In the first setup (refer to fig.\,\ref{I_AB(1)}), for $t_M+|b|<t<x_P+|a|$, we find descrambling behavior $I_{AB}>I^{\text{mix}}_{AB}$. As depicted in fig.\,\ref{fig:NSCR}, the left-moving mode of the local operator excitation stays in subsystem $A$, while the right-moving mode can spread over to subsystem $B$ in CFT$_2$. Therefore $S_{AB}$ gets suppressed and $I_{AB}$ increases. On the other hand, for $t>x_P+|a|$, the left-moving mode leaves subsystem $A$ and thus $S_A$ also decreases. Since the local excitation tends to break the quantum entanglement created by the localized TFD state, the standard scrambling behavior follows. The fact that the time at which descrambling switches to scrambling coincides with the time at which the left-moving mode of the local operator exits subsystem $A$ indicates that the left-moving mode in CFT\textsubscript{1} is entangled with other modes through the wormhole.

In the second setup (refer to fig.\,\ref{I_AB(2)}), at early times $t<|a|+x_P$, the right-moving mode of the local operator excitation transmits to subsystem $B$ and this adds to the correlation between $A$ and $B$. Thus 
$\Delta I_{AB}$ takes a very small but positive value. For intermediate times, $x_P+|a|<t<t_M+|a|$, we observe $I_{AB}<I^{\text{mix}}_{AB}$, i.e. the standard scrambling behavior. In this case, the left-moving mode of the local operator excitation leaves subsystem $A$, while the right-moving mode is still inside subsystem $A$ and spreads over to CFT$_2$, as illustrated in the left panel of fig.\,\ref{fig:SCR}. Since $S_A$ is suppressed compared to $S_{AB}$, the mutual information decreases. 

In the third setup (refer to fig.\,\ref{I_AB(3)}), for $t_M+|a|<t<t_M+|c|$, we find $I_{AB}<I^{\text{mix}}_{AB}$, which is the standard scrambling effect. In this case, as illustrated in the right panel of fig.\,\ref{fig:SCR}, the left-moving mode is outside of subsystem $A$, while the right-moving mode is inside subsystem $A$. Since the right-moving mode spreads over to subsystem $B$, again $S_A$ is highly suppressed, though $S_{AB}$ is not affected much. Therefore the mutual information gets reduced.

\section{Conclusion}\label{Sec:Concl}
In this paper, we studied how energy and quantum entanglement are transferred when two identical CFTs, denoted by CFT$_1$ and CFT$_2$, are locally entangled with each other. For this we considered the localized thermofield double (TFD) state (\ref{TFDstate}). In this state, the two CFTs are entangled at effective temperature $T_{\text{eff}}=1/\beta$, localized only around a small region of size $s$ at a specific time. In the AdS$_3/$CFT$_2$, it is dual to a localized wormhole in AdS$_3$, which has non-trivial time evolution as the wormhole travels at the speed of light.

We first calculated the profile of energy density due to the introduction of a local operator excitation (localized in a region of size $\delta$) and saw that holographic CFTs and the free scalar CFT share similar qualitative behavior, though the free CFT exhibits richer structure. The energy is transferred through the localized TFD state, or the wormhole in its gravitational dual. We found that as the ratio $\delta/s$ gets larger the amount of energy transmission gets enhanced. 

We also pointed out that the mechanism for the transmission of local excitation from CFT$_1$ to CFT$_2$ is closely related to quantum teleportation. As the ratio $\delta/s$ gets larger, we find that the teleportation becomes more efficient, being consistent with the mentioned energy density calculations. 
It is important to note that in the gravitational dual, this transmission occurs in a way that goes beyond event horizons. The reason why this is possible is that the insertion of local operator is not a unitary process. As far as we got aware, our argument is the first which observed that the local operator insertion realizes the quantum teleportation.

Next, we computed the entanglement entropy and mutual information in the localized TFD state. In the absence of local excitations, its time evolution can simply be understood in terms of the propagation of entangled pairs created by the localized TFD state. In the presence of a local operator excitation, we demonstrated that quantum entanglement gets transferred from one CFT to the other.

In \cite{Shenker:2013pqa,Caputa:2015waa}, the setup of shockwaves being added to a TFD background was studied. Within that setup, it was shown that mutual information between two subsystems belonging to different asymptotic regions decreases in the presence of shockwaves. This scrambling effect frequently shows up in our setup.
However, we also observed that under special conditions, the opposite phenomenon can occur, namely the enhancement of mutual information. We gave an intuitive explanation of why this happens. It is also intriguing to note that this new effect is tied directly to the ability of the localized TFD state to transmit information between two CFTs.

Our analysis of holographic CFTs assumed a valid approximation of correlation functions on a torus by those on a plane, taking the saddle point of the partition function that is dual to the BTZ black hole; this was also employed in the treatment of the mixed-state quench in \cite{Doi:2025oma}. It is possible that the approximation is not valid when $\delta/s=O(1)$, due to the backreaction of the local operator. Therefore, it would be an interesting future problem to perform a careful large $c$ analysis of correlation functions on a torus in order to clarify this potential issue.


\section*{Acknowledgements}
We are grateful to Yoshifumi Nakata for useful discussions.
This work is supported by by MEXT KAKENHI Grant-in-Aid for Transformative Research Areas (A) through the ``Extreme Universe'' collaboration: Grant Number 21H05187. TT is also supported by Inamori Research Institute for Science and by JSPS Grant-in-Aid for Scientific Research (B) No.~25K01000. KD is supported by JSPS KAKENHI Grant Number JP24KJ1466. UN acknowledges support from the scholarship of the Kyoto iUP Program.

\appendix

\section{Field-theoretic derivation of $S_{AB}$}\label{Append:Derivation}
When the subsystem is taken as a double interval $A \cup B$, the only new ingredient is entanglement entropy $S_{AB}$. As before, we first calculate the entanglement entropy evaluated from the wormhole background $\Delta S_{AB}$, which is calculated by taking the $n\to 1$ limit of 
\begin{align}
    \Delta S_{AB}^{(n)}
    &=\frac{1}{1-n}\log
    \left\{
    \frac{\langle O^{\dagger n}(\zeta_{O^\dagger})
    \,\sigma_1(\zeta_{\sigma_1})
    \,\bar{\sigma}_1(\zeta_{\bar{\sigma}_1})
    \,\sigma_2(\zeta_{\sigma_2})
    \,\bar{\sigma}_2(\zeta_{\bar{\sigma}_2})
    \,O^n(\zeta_O)\rangle
    }
    {\langle O^{\dagger}(\zeta_{O^\dagger})\,O(\zeta_O)\rangle^n
    \,\langle \sigma_1(\zeta_{\sigma_1})
    \, \bar{\sigma}_1(\zeta_{\bar{\sigma}_1})\,\sigma_2(\zeta_{\sigma_2})
    \,\bar{\sigma}_2(\zeta_{\bar{\sigma}_2})\rangle
    }
    \right\} \notag\\[5pt]
    &=\frac{1}{1-n}\log
    \left\{
    \frac{\langle O^{\dagger n}(\infty)
    \,\sigma_1(z_{\sigma_1})
    \,\bar{\sigma}_1(z_{\bar{\sigma}_1})
    \,\sigma_2(0)
    \,\bar{\sigma}_2(z)
    \,O^n(1)\rangle
    }
    {\langle O^{\dagger}(\infty)\,O(1)\rangle^n
    \,\langle \sigma_1(z_{\sigma_1})
    \, \bar{\sigma}_1(z_{\bar{\sigma}_1})\,\sigma_2(z_{\sigma_2})
    \,\bar{\sigma}_2(z_{\bar{\sigma}_2})\rangle
    }
    \right\}
\end{align}
where we have used the conformal map
\begin{equation}
    z(\zeta)=
    \frac{(\zeta_O-\zeta_{O^\dagger})(\zeta_{\sigma_2}-\zeta)}
    {(\zeta_O-\zeta_{\sigma_2})(\zeta_{O^\dagger}-\zeta)}
\end{equation}
with 
\begin{equation}
    z\coloneqq\frac{(\zeta_O-\zeta_{O^\dagger})
    (\zeta_{\sigma_2}-\zeta_{\bar{\sigma}_2})}
    {(\zeta_O-\zeta_{\sigma_2})
    (\zeta_{O^\dagger}-\zeta_{\bar{\sigma}_2})}
\end{equation}
It is known that to calculate
\begin{equation}
    G(z,\bar{z})=\frac{\langle O^{\dagger n}(\infty)
    \,\sigma_1(z_{\sigma_1})
    \,\bar{\sigma}_1(z_{\bar{\sigma}_1})
    \,\sigma_2(0)
    \,\bar{\sigma}_2(z)
    \,O^n(1)\rangle
    }
    {\langle O^{\dagger}(\infty)\,O(1)\rangle^n}
\end{equation}
it is sufficient to approximate the numerator of $G(z,\bar{z})$ by the factorization $\langle O^{\dagger n}\sigma_1\bar{\sigma}_1O^n\rangle\langle O^{\dagger n}\sigma_2\bar{\sigma}_2O^n\rangle$ or $\langle O^{\dagger n}\sigma_1\bar{\sigma}_2O^n\rangle\langle O^{\dagger n}\sigma_2\bar{\sigma}_1O^n\rangle$ in holographic CFTs where the central charge is taken to be sufficiently large \cite{Banerjee:2016qca}. As in the wormhole case, we call the first factorization disconnected phase and the second connected phase. This can be understood in the OPE language. The first factorization comes from the channel where the twist operators are first contracted in respective CFTs then contacted to the local operators, and the second factorization comes from the channel where $\sigma_1$ and $\sigma_2$ are first contracted to $\bar{\sigma}_2$ and $\bar{\sigma}_1$ and then get into the contraction with the local operators. 

In the disconnected phase, 
\begin{equation}
    G(z,\bar{z})=
    |\mathcal{F}(\infty,0,z,1;0,h_{O^n},h_\sigma)|^2
    |\mathcal{F}(\infty,z_{\sigma_1},z_{\bar{\sigma}_1},1;0,h_{O^n},h_\sigma)|^2
\end{equation}
with the conformal block
\begin{equation}
    \mathcal{F}(\infty,x_i,x_j,1;0,h_{O^n},h_\sigma)
    =\left(\frac{\alpha_H}{(1-x_i)^{\alpha_H}-(1-x_j)^{\alpha_H}}\right)^{2nh_\sigma}
    [(1-x_i)(1-x_j)]^{nh_\sigma(\alpha_H-1)}
\end{equation}
Then we have
\begin{align}
    \Delta S_{AB}^{(n)}
    =&\frac{1}{1-n}\log\left\{
    |z|^{4h_\sigma}|z_{\sigma_1}-z_{\bar{\sigma}_1}|^{4h_\sigma}
    \left|\frac{\alpha_H}{1-(1-z)^{\alpha_H}}\right|^{4nh_\sigma}
    |1-z|^{2nh_\sigma(\alpha_H-1)} 
    \right. \notag\\
    &\qquad\qquad\times \left.
    \left|\frac{\alpha_H}{(1-z_{\sigma_1})^{\alpha_H}-(1-z_{\bar{\sigma}_1})^{\alpha_H}}\right|^{4nh_\sigma}
    |(1-z_{\sigma_1})(1-z_{\bar{\sigma}_1})|^{2nh_\sigma(\alpha_H-1)}\right\}
\end{align}
and thus
\begin{align}
    \Delta S_{AB}&=\frac{c}{6}\log\left\{
    \frac{1}{|z|^2|z_{\sigma_1}-z_{\bar{\sigma}_1}|^2}
    \left|\frac{1-(1-z)^{\alpha_H}}{\alpha_H}\right|^2
    |1-z|^{1-\alpha_H} 
    \right.   \notag\\
    &\qquad\qquad\times \left.
    \left|\frac{(1-z_{\sigma_1})^{\alpha_H}-(1-z_{\bar{\sigma}_1})^{\alpha_H}}{\alpha_H}\right|^2
    |(1-z_{\sigma_1})(1-z_{\bar{\sigma}_1})|^{1-\alpha_H}\right\}\notag\\
    &=\Delta S_A +\Delta S_B
\end{align}
where we have noted that 
\begin{equation}
    \frac{1-z_{\bar{\sigma}_1}}{1-z_{\sigma_1}}
    =1-\frac{(\zeta_O-\zeta_{O^\dagger})(\zeta_{\sigma_1}-\zeta_{\bar{\sigma}_1})}{(\zeta_O-\zeta_{\sigma_1})(\zeta_{O^\dagger}-\zeta_{\bar{\sigma}_1})}\eqqcolon1-z'
\end{equation}
On the other hand, for the connected phase, we have
\begin{equation}
    G(z,\bar{z})=
    |\mathcal{F}(\infty,z_{\sigma_1},z,1;0,h_{O^n},h_\sigma)|^2
    |\mathcal{F}(\infty,0,z_{\bar{\sigma}_1},1;0,h_{O^n},h_\sigma)|^2
\end{equation}
and thus
\begin{align}
    \Delta S_{AB}&=\frac{c}{6}\log\left\{
    \frac{1}{|z_{\bar{\sigma}_1}|^2|z-z_{\sigma_1}|^2}
    \left|\frac{1-(1-z_{\bar{\sigma}_1})^{\alpha_H}}{\alpha_H}\right|^2
    |1-z_{\bar{\sigma}_1}|^{1-\alpha_H} 
    \right.   \notag\\
    &\quad\quad\times \left.
    \left|\frac{(1-z_{\sigma_1})^{\alpha_H}-(1-z)^{\alpha_H}}{\alpha_H}\right|^2
    |(1-z_{\sigma_1})(1-z)|^{1-\alpha_H}\right\}
\end{align}
Define $\theta_1,\theta_2$ by 
\begin{align}
    &1-z_{\sigma_1}=e^{i\theta_1}\,,\quad
    1-\bar{z}_{\sigma_1}=e^{-i\bar{\theta}_1}  \\
    &1-z_{\bar{\sigma}_1}=e^{i\theta_2}\,,\quad
    1-\bar{z}_{\bar{\sigma}_1}=e^{-i\bar{\theta}_2}
\end{align}
then $\Delta S_{AB}$ can also be written as 
\begin{equation}
    \Delta S_{AB}=\left\{
    \begin{aligned}
        &\frac{c}{6}\log\left\{
        \frac{1}{\alpha_H^4}
        \frac{\sin\frac{\alpha_H\theta}{2}\sin\frac{\alpha_H\bar{\theta}}{2}}{\sin\frac{\theta}{2}\sin\frac{\bar{\theta}}{2}}
        \frac{\sin\frac{\alpha_H}{2}(\theta_1-\theta_2)\sin\frac{\alpha_H}{2}(\bar{\theta}_1-\bar{\theta}_2)}{\sin\frac{\theta_1-\theta_2}{2}\sin\frac{\bar{\theta}_1-\bar{\theta}_2}{2}} 
        \right\} \,,\quad\text{dis} \\[8pt]
        &\frac{c}{6}\log\left\{
        \frac{1}{\alpha_H^4}
        \frac{\sin\frac{\alpha_H\theta_2}{2}\sin\frac{\alpha_H\bar{\theta}_2}{2}}{\sin\frac{\theta_2}{2}\sin\frac{\bar{\theta}_2}{2}}
        \frac{\sin\frac{\alpha_H}{2}(\theta_1-\theta)\sin\frac{\alpha_H}{2}(\bar{\theta}_1-\bar{\theta})}{\sin\frac{\theta_1-\theta}{2}\sin\frac{\bar{\theta}_1-\bar{\theta}}{2}} 
        \right\} \,,\quad\text{con}
    \end{aligned}
    \right.
\end{equation}
where 'dis' and 'con' denote the disconnected and connected phases respectively.
To obtain the correct entanglement entropy with the insertion of the local operator, we need to sum up $\Delta S$ and $S^\text{mix}$ for disconnected phase and connected phase respectively and then pick up the smaller one. Therefore, the entanglement entropy for subsystem $AB$ is given by 
\begin{equation}\label{APS_AB}
    S_{AB}=\underset{\text{con,\,dis}}{\min}\big\{
    \Delta S_{AB}^{{\text{dis}}}+S^{\text{mix},\text{dis}}_{AB},
    \Delta S_{AB}^{\text{con}}+S^{\text{mix},\text{con}}_{AB}
    \big\}
\end{equation}

\clearpage
\bibliographystyle{JHEP}
\bibliography{WHEnt}


\end{document}